\documentstyle[12pt,aps]{revtex}  

\begin{document}
\title{Quantum spinor field in the FRW universe with a constant electromagnetic
background}
\author{S.P. Gavrilov\thanks{%
On leave from Tomsk Pedagogical University, 634041 Tomsk, Russia; present
address: Departamento de F\'\i sica, CCET, Universidade Federal de Sergipe,
49000-000 Aracaju, SE, Brazil; e-mail: gavrilov@sergipe.ufs.br}, D.M. Gitman%
\thanks{%
Instituto de F\'{\i }sica, Universidade de S\~{a}o Paulo, Caixa Postal
66318, 05315-970-S\~{a}o Paulo, SP, Brazil; e-mail: gitman@fma.if.usp.br}
and A.E. Gon\c{c}alves \thanks{%
Dept. F\'\i sica, CCE, Universidade Estadual de Londrina, CP 6001. CEP
86051-990, Londrina, PR, Brasil; e-mail: goncalve@fisica.uel.br}}
\date{\today }
\maketitle

\begin{abstract}
The article is a natural continuation of our paper 
{\em Quantum scalar field in FRW Universe with constant electromagnetic 
background}, Int. J. Mod. Phys. {\bf A12}, 4837 (1997). We 
generalize the latter 
consideration to the case of massive spinor field, which is  placed in
FRW  Universe of special type with a
constant electromagnetic field. To this end special sets of exact 
solutions of Dirac equation in the background under consideration 
are constructed and classified. Using these solutions  
representations for out-in, in-in, and out-out spinor Green
functions  are explicitly constructed as proper-time integrals over
the corresponding contours in complex proper-time plane.
The vacuum-to-vacuum transition
amplitude and number of created  particles are found and vacuum
instability is discussed. The mean values of the current and
energy-momentum tensor  are evaluated, and different
approximations for them are presented. The back reaction related to 
 particle creation and to the polarization of the unstable vacuum is 
estimated in different regimes.

\end{abstract}

\draft

\section{Introduction}

The article is a natural continuation of our paper \cite{GGO}.
 Here we generalize the latter 
consideration to the case of a massive spinor field which is  placed in
Friedmann-Robertson-Walker (FRW)  Universe of special type with
constant electromagnetic field. First, we ought to repeat briefly a 
motivation for such an activity.

It is quite well known fact that quantum field theory in an external
background is, generally speaking, theory with unstable vacuum. The
vacuum instability leads to many interesting features, among which
particle creation from the vacuum is one of the
most beautiful non-perturbative phenomenon. 
One has to treat it exactly in regard
to the external field.
The latter has been realized
long ago by Schwinger \cite{Sch1}.
The particle creation effects together with back reaction issue are
important  in black hole physics and in dynamics of the early Universe (EU)
(see, for example, \cite{AM,TWo} and references  therein).

In  quantum field theory with unstable vacuum it
is necessary to construct different kinds of Green functions (GF), 
e.g. besides 
the causal GF (out-in GF) one has to use so called in-in GF, out-out
GF, and so on \cite{BdW,GG1977,FG} (for a review and technical details see
\cite{FGS}). General methods of such GF construction in electromagnetic
 background have been  developed in \cite{GG1977,FG}. A possible 
generalization of the formalism to an external gravitational
background has been given in Ref. \cite{BFG}. Since Ref.
\cite{Sch1} it has been known that causal (out-in) GF may be
presented as a proper-time integral over a real infinite
contour. At the same time, in the unstable vacuum  case
the in-in
and out-out GF differ from the causal one. It was shown
\cite{GGS,GGG} there are examples of external fileds (electromagnetic
with constant uniform invariants) when
 these functions may be presented by the same
proper-time integrals (with the same integrand) but over another
contours in the complex proper-time plane. Then, it is not difficult
to compare contributions from the in-in GF and from the causal one.
The complete set of GF
mentioned is necessary for the construction of the 
Furry picture in interacting 
theories, and even in noninteracting cases one has to use them to define, for
example, the back reaction of particles created and to construct
different kinds of effective actions  (EA) (for a general 
introduction to EA in background field method, see \cite{BOS},
and for review of modern generalizations, see \cite{G}).
The such proper-time representation of GF may be the necessary step in
the study of chiral symmetry breaking in QED and the four-fermion
models under the action of gravitational and electromagnetic fields
(see \cite{IMO} and references therein).
>From another point, in-in GF which gives the origin to
in-in EA maybe used in more realistic theories,
like GUT theories with scalars, spinors and vectors in order 
to analyse the properties of above electro-gravitational background
 in the EU. For example, one of extremely interesting questions 
there is: can we realise the asymptotic conformal invariance phenomenon
(which means that theory becomes approximately conformally
invariant  at large curvature) \cite{BO2} even for in-in EA,
or in other words for mean values in EU. Taking into
account our recent study of in-in GF  structure for scalars \cite
{GGO} it looks quite interesting next application of above calculation.

It may be likely that EU  is filled with some type of
electromagnetic fields. For example, recently (see \cite{TW,GGV} and 
references therein) the possibility of existence and role of
primordial magnetic fields in EU have been discussed. From
another point the possibility of existence of electromagnetic field in
the EU has been discussed long ago in \cite{SD,BO1}. It
has been shown there that the presence of the electrical field in the 
EU increases
significantly  the gravitational particle creation from the
vacuum.
 In principle, this process may be considered as a source for
the dominant part of the Universe mass. 

Bearing in mind the above cosmological motivations it is becoming
interesting to study the quantum field theory in curved background
with electromagnetic field (of a special form in order to solve the
problem analytically). In the present paper we are going to consider a
massive spinor field placed in the expanding FRW
Universe with the scale factor $\Omega (\eta)$ (in terms of the
conformal time) $\Omega^2 (\eta)=b^2\eta^2+a^2$. Such a scale factor
corresponds to the expanding radiation-dominated FRW Universe. In
terms of physical time $t$ the corresponding metric may be written as
follows:
\begin{equation}\label{1}
ds^2=dt^2-\Omega^2(t)(dx^2+dy^2+dz^2)\;,
\end{equation}
where for small times $|t|\ll a^2/b,\; \Omega^2(t)\simeq
a^2[1+(bt/a^2)^2]$, and for large times $|t|\gg a^2/b,\; \Omega^2(t)\simeq
2b|t|$ (see \cite{SD}). Moreover, such FRW Universe will be filled by
the constant electromagnetic field. 

 Thus, we start from the  theory of massive spinor in above
background.  Making a conformal transformation 
 we remain with QED in flat background 
but with time-dependent mass
(QED-$\Omega $ theory).
In the Sect.II  special sets of exact 
solutions of Dirac equation in QED-$\Omega $ theory
are constructed and classified as corresponding to particles and antiparticles 
at $t\rightarrow \pm \infty$. In the Sect.III,
using these solutions,  
representations for out-in, in-in, and out-out spinor GF
  are explicitly constructed as  
proper-time integrals over
the corresponding  contours in complex proper-time plane.
 As far as we know,
 it is a first explicit example for the proper-time representations for
complete set of spinor  GF in gravitational-electromagnetic
background.
In the Sect.IV we are interested in to reveal global features of the theory.
The vacuum-to-vacuum transition amplitudes 
and number of created  particles are found and vacuum
instability is discussed. It is seen the creation process is a
coherent effect of both fields. 
The all  mean values of the current and
energy-momentum tensor  are presented in the same manner as the
proper-time integrals, and 
evaluated. The different
approximations for them are investigated. 
The back reaction  produced by both of particles created
from a vacuum and polarization of an unstable vacuum 
estimated in different regimes. It is shown a behaviour of such
components in time are quite different.

\section{Classified sets of exact solutions}

In this Section we study exact solutions of the Dirac equation in an
external constant uniform electromagnetic background and in a time-dependent
mass-like potential, which effectively reproduces effects of a gravitational
background (solutions of the Dirac equation of QED-$\Omega $ theory),

\begin{eqnarray}\label{H1}
&&({\cal P}_\mu \gamma ^\mu -M\,\Omega )\psi \left( x\right) =0\, ,\;\;  
\Omega =\Omega (x_0)=\sqrt{a^2+b^2x_0^2}\;, \\
&&{\cal P}_\mu =i\partial _\mu
-qA_\mu \left( x\right)\;,\;\;
\left[ \gamma ^\mu ,\gamma ^\nu \right] _{+}=2\eta ^{\mu \nu },\quad \eta
^{\mu \nu }={\rm diag}\left( 1,-1,-1,-1\right) ,\nonumber
\end{eqnarray}
where $ x^0=\eta$ is conformal time,
$q$ is charge of a particle, for example,$\;q=-|e|\,$ for electron. 
The time-independent inner product of
the solutions of the equation (\ref{H1}) may be chosen as
\begin{equation}
(\psi ,\psi ^{\prime })=\int \bar{\psi}(x)\gamma ^0\psi ^{\prime }(x)d{\bf x}%
.  \label{w1}
\end{equation}
As usual, it is convenient to present $\psi (x)$ in the following form 
\begin{equation}
\psi (x)=\left( {\cal P}_\mu \gamma ^\mu +M\Omega \right) \phi (x)\;.
\label{H2}
\end{equation}
Then the functions $\phi $ have to obey the squared Dirac equation, 
\begin{eqnarray}
&&\left( {\cal P}^2-\left( M\Omega \right) ^2-\frac q2\sigma ^{\mu \nu
}F_{\mu \nu }+iM\partial _0\Omega \gamma ^0\right) \phi (x)=0\;,\;\;
\label{H3a} \\
&&\;F_{\mu \nu }=\partial _\mu A_\nu (x)-\partial _\nu A_\mu (x)\;,\;\sigma
^{\mu \nu }=\frac i2[\gamma ^\mu ,\gamma ^\nu ]\;.  \nonumber
\end{eqnarray}
The external electromagnetic field in our case consists of a
constant uniform electric ($E$) and parallel to it magnetic ($H$) fields, 
\begin{equation}
F_{03}=E,\;\;F_{\mu \nu }^{\perp }=H\left( \delta _\mu ^2\delta _\nu
^1-\delta _\nu ^2\delta _\mu ^1\right) .  \label{a2}
\end{equation}
For such a field we select the following potentials: 
\begin{equation}
A_0=0,\;A_3=Ex^0,\;A_i=A_i^{\perp }=-H\,x_2\,\delta _i^1,\;\;i=1,2.
\label{a4}
\end{equation}
For 
$\,b=0$ we have a usual flat-space case with the mass $m=aM$. In this case
particle-antiparticle classified solutions of the equation and
all the GF
were found in \cite{Nik,GGS,GGG}. The case $\,b\neq 0\,$
is of special interest for us. In the case one can consider spinor field in
the conformally-flat Universe (with scale factor $\,\Omega $)$\,$ filled by
a constant uniform electromagnetic field. Making a standard conformal
transformation of the gravitational metric and spinor field, we arrive to a
theory in flat space-time with a time-dependent mass (QED-$\Omega $). The
corresponding field equation is given by (\ref{H1}) (an electromagnetic
field should not be transformed under the conformal transformation). Note
that the such a conformal transformation may be used also for interacting
theories \cite{BOS,BO2}. Thus, $\,$ Eq. (\ref{H1}) is actually relevant
 to
the quantum spinor field in the expanding FRW Universe with the external
constant \
electromagnetic\ field. 

Remember \cite{SD} that the
expansion law 
with $a\neq 0$ is necessarily connected with a nonzero energy
density of the cosmological substratum in the early,
radiation-dominated phase of the Universe. The minimum value of $\Omega $ ($%
\Omega _{min}=a$) is caused by the strong interaction. 
It is difficult to find
solutions of the equation (\ref{H3a}) with the last
spin term. Nevertheless, in many interesting cases with strong background
part of this term 
(for region of small $(x_0)^2\lesssim (a/b)^2$) may be treated
perturbatively starting from the reduced
 squared Dirac equation: 
\begin{equation}
\left[ {\cal P}^2-\left( M\Omega \right) ^2-\frac q2\sigma ^{\mu \nu }F_{\mu
\nu }+ibM\gamma ^0\right] \phi (x)=0\;,\;\;  \label{H3b}
\end{equation}
where the asymptotic form of the last spin term is used ($\partial _0\Omega
\rightarrow b$ for large enough  $x_0$). If the intensity of an electric
field is more than
 the characterizing parameter of a mass-like potential, $%
(qE)^2\gg (bM)^2$, or its potential is not intense, $a^2M/b\gg 1$, then both
the last spin term of equation (\ref{H3a}) and of the 
equation (\ref{H3b}) can be
disregarded. 
 If $(qE)^2\lesssim (bM)^2,$ one can not neglect
the contribution from the  term $iM\partial _0\Omega \gamma ^0$. 
However, our main interest is to study an
intense gravitational background when $a^2M/b\ll 1$. Then, as it may be seen
from explicit solutions of the equation (\ref{H3b}), one can calculate any
spin contributions to matrix elements of operators using solutions
of equation (\ref{H3b}) with a relative accuracy of the order of $a^2M/b.$
And one may use these sets as a basis to construct solutions of the
equation (\ref{H3a}) perturbatively, considering the term $iM(\partial
_0\Omega -b)\gamma ^0$ as a perturbation. Moreover,  the large
time ($(x_0)^2\gg (a/b)^2$ ) asymptotic solutions  of both  equations (%
\ref{H3a}) and (\ref{H3b}) are the same.

To construct the above mentioned generalized Furry picture for both  QED
and QED-$\Omega $  one has to find special sets of
classified  solutions of the equation (\ref{H1}), namely, two complete
and orthonormal sets of solution: $\left\{ _{\pm }\psi _{\left\{ n\right\}
}(x)\right\} $, which describes particles (+) and antiparticles ($-$) in the
initial time instant ($x^0\rightarrow -\infty )$, and $\left\{ ^{\pm }\psi
_{\left\{ n\right\} }(x)\right\} $, which describes particles (+) and
antiparticles ($-$) in the final time instant ($x^0\rightarrow +\infty ).$
According to the general approach \cite{GG1977} , which can be easily
adapted to QED-$\Omega $, such solutions obey the following asymptotic
conditions 
\begin{eqnarray}
\ &&H_{o.p.}(x^0)\;{}_\zeta \psi _{\left\{ n\right\} }(x)={}_\zeta
\varepsilon \;{}_\zeta \psi _{\left\{ n\right\} }(x),\;\;\;,{\rm sgn}%
\;{}_\zeta \varepsilon =\zeta ,\;x^0\rightarrow -\infty \;,  \nonumber
\label{e8} \\
\ &&H_{o.p.}(x^0)\;{}^\zeta \psi _{\left\{ l\right\} }(x)={}^\zeta
\varepsilon \;{}^\zeta \psi _{\left\{ l\right\} }(x),\;{\rm sgn}\;{}^\zeta
\varepsilon =\zeta ,\;\;x^0\rightarrow +\infty \;, \;\;\zeta=\pm\;,
 \label{w2}
\end{eqnarray}
where $\zeta ,\left\{ n\right\} $ and $\zeta ,\left\{ l\right\} $ are 
complete sets of quantum numbers which characterize solutions $_\zeta \psi
_{\left\{ n\right\} }(x)$ and $^\zeta \psi _{\left\{ l\right\} }(x)$
respectively, $H_{o.p.}\left( x^0\right) =\gamma ^0(M\Omega -\gamma ^i{\cal P%
}_i)$ is  one-particle Dirac Hamiltonian;
 $^{+}\varepsilon $, $_{+}\varepsilon $ are particle
quasi-energies and $|^{-}\varepsilon |$ and $|_{-}\varepsilon |$ are
antiparticles quasi-energies. All the information about the processes of
particles scattering and creation by an external field (in zeroth order with
respect to the radiative corrections) can be extracted from the
decomposition coefficients, which form the matrices
  $G\left( {}_\zeta |{}^{\zeta ^{\prime
}}\right) $, 
\begin{equation}
{}^\zeta \psi (x)={}_{+}\psi (x)G\left( {}_{+}|{}^\zeta \right) +{}_{-}\psi
(x)G\left( {}_{-}|{}^\zeta \right) \;.  \label{we10}
\end{equation}
The matrices $G\left( {}_\zeta |{}^{\zeta ^{\prime }}\right) $ obey the
following relations, 
\begin{eqnarray}
\ &&G\left( {}_\zeta |{}^{+}\right) G\left( {}_\zeta |{}^{+}\right)
^{\dagger }+G\left( {}_\zeta |{}^{-}\right) G\left( {}_\zeta |{}^{-}\right)
^{\dagger }={\bf I},  \nonumber  \label{e10a} \\
\ &&G\left( {}_{+}|{}^{+}\right) G\left( {}_{-}|{}^{+}\right) ^{\dagger
}+G\left( {}_{+}|{}^{-}\right) G\left( {}_{-}|{}^{-}\right) ^{\dagger }=0\;,
\label{w15}
\end{eqnarray}
where ${\bf I}$ is the identity matrix. All GF in the formalism may be
also constructed with the sets  $\left\{ _{\pm }\psi _{\left\{
n\right\} }(x)\right\} $ and $\left\{ ^{\pm }\psi _{\left\{ n\right\}
}(x)\right\} $.
Below we are going to present such solutions. 

The
functions $\phi (x)\,$ can be written in the following form: 
\begin{equation}
\phi _{p_3p_1n\xi r}(x)=\phi _{p_3n\xi r}(x_{\parallel })\,\phi
_{p_1nr}(x_{\perp })v_{\xi r}\quad ,  \label{a6}
\end{equation}
where 
$\,x_{\perp }^{\mu}=(0,x^1,x^2,0),\quad x_{\parallel }^\mu
=(x^0,0,0,x^3) $; $\left\{ p_3,p_1,n,\xi ,r\right\} $ is a complete set
of quantum numbers. Among them $p_3$ and $p_1$ are momenta of the continuous
spectrum, $n$ is an integer quantum number, $\xi =\pm 1$ and $r=\pm 1$ are spin
quantum numbers;
 $v_{\xi r}$ are some constant orthonormal spinors, $v_{\xi r}^{\dagger
}v_{\xi r^{\prime }}=\delta _{rr^{\prime }}.$ The eq.(\ref{H3b}) allows one
to subject these spinors to some supplementary conditions, 
\begin{eqnarray}
&& \Xi v_{\xi r}=\xi v_{\xi r},\;\Xi =\gamma ^0(qE\gamma ^3-bM)/\rho
,\smallskip \ \;\rho =\sqrt{(qE)^2+(bM)^2}\,,  \nonumber\\
&&Rv_{\xi r}=rv_{\xi r},\smallskip \ \ R=\mbox{sgn}\left( qH\right) i\gamma
^1\gamma ^2.  \label{w12}
\end{eqnarray}
If $H\neq 0,$ the function $\,\phi _{p_1nr}(x_{\perp })$ has the form
\[
\phi _{p_1nr}(x_{\perp })= 
\left( \frac{\sqrt{|qH|}}{2^{n+1}\pi ^{\frac 32}n!}\right) ^{1/2}\exp
\left\{ -ip_1x^1-\frac{X^2}{2}\right\} 
{\cal H}_n\left(X \right)
,\;\;
X= \sqrt{|qH|}\left( x^2+\frac{p^1}{qH}\right),
\]
where ${\cal H}_n(x)$ are Hermite polynomials with integer $n=0,1,\ldots $%
. If $H=0,$ the discrete quantum number $n$ has to be replaced by the
momentum $p_2,$ and the corresponding function has the form $\phi
_{p_1nr}(x_{\perp })=(2\pi )^{-1}\exp \left\{ -i\left( p_1x^1+p_2x^2\right)
\right\} .$ Let us present the function 
$\phi(x_{\parallel })$
as follows 
\begin{equation}
\phi _{p_3n\xi r}(x_{\parallel })\,=(2\pi )^{-1/2}\,e^{-ip_3x^3}\phi
_{p_3n\xi r}(x^0)\quad ,\;\; \phi _{p_3n\xi r}(x^0)=
\phi _{p_3n\xi r}(x^0,p_z)|_{p_z=0}\; \label{a11}
\end{equation}
where $\phi _{p_3n\xi r}(x^0,p_z)$ is a solution of equation 
\begin{equation}
\left[ \left( i\frac \partial {\partial \widetilde{\eta }}\right) ^2-\left(
p_z-\rho \widetilde{\eta }\right) ^2-\rho \lambda -i\rho \xi \right] \phi
_{p_3n\xi r}\left( x^0,p_z\right) =0\;,  \label{a12}
\end{equation}
with $\widetilde{\eta } =x^0-\,\rho ^{-2}qEp_3,\quad \rho 
\lambda =p_3^2(bM/\rho
)^2+\omega +a^2M^2\;,$
\[
\omega =\left\{ 
\begin{array}{ll}
|qH|(2n+1-r),\;n=0,1,\ldots \;\;, & H\neq 0 \\ 
p_1^2+p_2^2\;\;, & H=0
\end{array}
\right. \;.
\]
One can form  two complete sets $\left\{ _{\pm }\phi _{p_3n\xi r}\left(
x^0,p_z\right) \right\} $ and $\left\{ ^{\pm }\phi _{p_3n\xi r}\left(
x^0,p_z\right) \right\} $ of the solutions of equation (\ref{a12}) using
the functions 
\begin{eqnarray}
&&\ {_{+}^{-}}\phi _{p_3n\xi r}\left( x^0,p_z\right) =C_\xi D_{\nu -\xi
/2}[\pm (1-i)\tau ],\; \tau =\frac 1{\sqrt{\rho }}
\left( \rho \widetilde{\eta }-p_z\right), \nonumber \\
\; &&{_{-}^{+}}\phi _{p_3n\xi r}\left( x^0,p_z\right) =C_\xi ^{\prime
}D_{-\nu -1+\xi /2}\left[ \pm (1+i)\tau \right],\;
\nu =\frac{i\lambda }2-\frac 12\;. \label{a13}
\end{eqnarray}
Similar  solutions were first presented in \cite{Nik}. Then,
solutions of equation (\ref{H3b}) $\phi (x)\,$ can be constructed as follows,

\begin{eqnarray}
_{\pm }\phi _{p_3p_1n\xi r}(x) &=&\smallskip \ _{\pm }\phi _{p_3p_1n\xi
r}(x,p_z)|_{p_z=0},\;  \label{H5} \\
\smallskip \ _{\pm }\phi _{p_3p_1n\xi r}(x,p_z) &=&(2\pi
)^{-1/2}\,e^{-ip_3x^3}\smallskip \ _{\pm }\phi _{p_3n\xi r}(x^0,p_z)\,\phi
_{p_1nr}(x_{\perp })v_{\xi r},  \nonumber
\end{eqnarray}
and in the same form with $\left( \pm \right) $ indices above.

One can verify that
the solutions of the Dirac equation with different $\xi $,
namely, $\left( {\cal P}_\mu \gamma ^\mu +M\Omega \right) {}_{\pm }\phi
_{p_3,p_1,n,+1,r}(x)$ and $\left( {\cal P}_\mu \gamma ^\mu +M\Omega \right)
{}_{\pm }\phi _{p_3,p_1,n,-1,r}(x),$ or $\left( {\cal P}_\mu \gamma ^\mu
+M\Omega \right) {}^{\pm }\phi _{p_3,p_1,n,+1,r}(x)$ and $\left( {\cal P}%
_\mu \gamma ^\mu +M\Omega \right) {}^{\pm }\phi _{p_3,p_1,n,-1,r}(x)$ are
linearly dependent for each sign ''$+$'' or ''$-$''. Thus, to construct the
complete sets we may use only the following sets of solutions: 
\begin{eqnarray}
{}\ {}{}_{\pm }\psi _{p_3p_1nr}(x) &=&\left( {\cal P}_\mu \gamma ^\mu
+M\Omega \right) \smallskip \ _{\pm }\phi _{p_3,p_1,n,+1,r}(x)\ {},
\label{w14} \\
{}{}^{\pm }\psi _{p_3p_1nr}(x) &=&\left( {\cal P}_\mu \gamma ^\mu +M\Omega
\right) \smallskip \ ^{\pm }\phi _{p_3,p_1,n,+1,r}(x)\;.\label{w14a}   
\end{eqnarray}
Choosing the coefficients $C$ and $C'$ in (\ref{a13}) as follows: 
$
C_{+1}=(2\rho )^{-1/2}\,\exp \left( -\pi \lambda /8\right) $ and
$\ C_{+1}^{\prime }=(\rho \lambda )^{-1/2}\,\exp \left( -\pi \lambda /8%
\right) ,  
$
one gets two complete sets $\left\{ \ {}{}_{\pm }\psi
_{p_3p_1nr}(x\right\} $ and $\left\{ {}{}^{\pm }\psi _{p_3p_1nr}(x)\right\} $
of orthonormalized solutions of the equation (\ref{H1}). These solutions are
classified as particles (+) and antiparticles ($-$) at $x^0\rightarrow \pm
\infty $ according to the asymptotic forms of the corresponding 
 quasienergies,
 ${}_\zeta \varepsilon =\zeta \rho |x^0|$ and ${}^\zeta
\varepsilon =\zeta \rho |x^0|$ ${}$(see \cite{GG1} for additional arguments
advocating such a classification). It matches with classification \cite{Nik}
of similar solutions in QED.

According to the above discussion the solutions (\ref{w14}) and (\ref{w14a})
of the Dirac equation may serve in an intense gravitational background, $%
a^2M/b\ll 1$, and in the other cases mentioned after Eq. (\ref{H3b}).
Satisfying the Cauchy conditions one can see the solutions (\ref{w14}) are
valid with $x^0<0$, $|x^0|\gg a/b$ and the solutions (\ref{w14a}) are valid
with $x^0>0$, $|x^0|\gg a/b$ for any intensity of the background.

To find the matrices $G\left( {}_\zeta |{}^{\zeta ^{\prime
}}\right) $  defined by (\ref{we10}%
), it is convenient to use an asymptotic form of the solutions. 
Using (\ref{H3b}) and (\ref{w12}) we get
\begin{equation}
G\left( {}_\zeta |{}^{\zeta ^{\prime }}\right) _{ll^{\prime }}=\delta
_{l,l^{\prime }}\;g\left( {}_\zeta |{}^{\zeta ^{\prime }}\right)
\;,\;\;l=(p_3,p_1,n,r)\;,\smallskip \ l^{\prime }=(p_3{}^{\prime
},p_1{}^{\prime },n^{\prime },r^{\prime })\;,  \label{e37a}
\end{equation}
where 
\begin{equation}
g(_\zeta |^{\zeta ^{\prime }})={}_\zeta \phi _{p_3,n,+1,r}^{*}\left(
x^0,p_z\right) \,i\stackrel{\leftrightarrow }{\partial }_0\left( i\partial
_0-\rho \tilde{\eta}\right) \smallskip \ {}^{\zeta ^{\prime }}\phi
_{p_3,n,+1,r}\left( x^0,p_z\right) \;.  \label{a15}
\end{equation}

\section{Green functions}

Let us start with out-in GF which is
 the causal propagator 
\begin{equation}
\ S^c(x,x^{\prime })=c_v^{-1}i<0,out|T\psi (x)\bar{\psi}(x^{\prime
})|0,in>,\;\;c_v=<0,out|0,in>\;.  \label{w23}
\end{equation}
Here $\psi (x)$ is quantum spinor field 
 satisfying the Dirac equation (\ref{H1}), $|0,in>$ and $%
|0,out>$ are the initial and the final vacuum, and $%
c_v$ is the vacuum-to-vacuum transition amplitude. The propagator $\
S^c(x,x^{\prime })$ obeys the equation 
\begin{equation}
\left( {\cal P}_\mu \gamma ^\mu -M\Omega \right) \ S^c(x,x^{\prime
})=-\delta ^{(4)}(x-x^{\prime })\;.  \label{w24}
\end{equation}
Another important singular
function is the commutation function 
$
S(x,x^{\prime })=i\left[ \psi (x),\bar{\psi}(x^{\prime })\right] _{+}.
$
It obeys the homogeneous Dirac equation (\ref{H1}) and the initial condition 
$
\left. S(x,x^{\prime })\right| _{x_0=x_0^{\prime }}=i\gamma ^0\delta ({\bf x}%
-{\bf x}^{\prime }).  
$
Besides,  the following GF are studied
 \cite{GG1977,FG,FGS}: 
\begin{eqnarray}
&&S_{in}^c(x,x^{\prime }) =i<0,in|T\psi (x)\bar{\psi}(x^{\prime
  })|0,in>, \;\;
S_{in}^{\bar{c}}(x,x^{\prime }) =i<0,in|\psi (x)\bar{\psi}(x^{\prime
})T|0,in>,
\nonumber \\
&&S_{in}^{-}(x,x^{\prime }) =i<0,in|\psi (x)\bar{\psi}(x^{\prime })|0,in>,\;\; 
S_{in}^{+}(x,x^{\prime }) =i<0,in|\bar{\psi}(x^{\prime })\psi (x)|0,in>, \;
  \nonumber \\
&&S_{out}^c(x,x^{\prime }) =i<0,out|T\psi (x)\bar{\psi}(x^{\prime })|0,out>.
\label{w25}
\end{eqnarray}
Here the $T$-product acts on both sides: it orders the field
operators to the right of its and antiorders them to the left. The functions $%
S_{in}^c$ and 
 $S_{out}^c$ obey the equation (\ref
{w24}), $S^{\mp }$ satisfy the equation (\ref{H1}), and $%
S_{in}^{\bar{c}}$ obeys the equation 
$
\left( {\cal P}_\mu \gamma ^\mu -M\Omega \right) \ S_{in}^{\bar{c}%
}(x,x^{\prime })=\delta ^{(4)}(x-x^{\prime })\;.  
$

One can express the
GF via the solutions (\ref{w14}) and (\ref{w14a}) \cite
{GG1977,FG,FGS}: 
\begin{eqnarray}
S^c\left( x,x^{\prime }\right) &=&\theta \left( x_0-x_0^{\prime }\right)
S^{-}\left( x,x^{\prime }\right) -\theta \left( x_0^{\prime }-x_0\right)
S^{+}\left( x,x^{\prime }\right) ,  \nonumber \\
S\left( x,x^{\prime }\right) &=&S^{-}\left( x,x^{\prime }\right)
+S^{+}\left( x,x^{\prime }\right) ,  \nonumber\\
S_{in}^c\left( x,x^{\prime }\right) &=&\theta \left( x_0-x_0^{\prime
}\right) S_{in}^{-}\left( x,x^{\prime }\right) -\theta \left( x_0^{\prime
}-x_0\right) S_{in}^{+}\left( x,x^{\prime }\right) ,  \nonumber \\
S_{in}^{\bar{c}}\left( x,x^{\prime }\right) &=&\theta \left( x_0^{\prime
}-x_0\right) S_{in}^{-}\left( x,x^{\prime }\right) -\theta \left(
x_0-x_0^{\prime }\right) S_{in}^{+}\left( x,x^{\prime }\right) ,
\nonumber\\
S_{out}^c\left( x,x^{\prime }\right) &=&\theta \left( x_0-x_0^{\prime
}\right) S_{out}^{-}\left( x,x^{\prime }\right) -\theta \left( x_0^{\prime
}-x_0\right) S_{out}^{+}\left( x,x^{\prime }\right) ,  \label{w33} 
\end{eqnarray}
where
\begin{eqnarray}
S^{-}\left( x,x^{\prime }\right) &=&i\int_{-\infty }^{+\infty
}dp_3dp_1\sum_{nr}{}\smallskip \ ^{+}\psi _{p_3p_1nr}(x)g\left( \left.
_{+}\right| ^{+}\right) ^{-1}\smallskip \ _{+}\bar{\psi}_{p_3p_1nr}\left(
x^{\prime }\right) ,  \nonumber \\
S^{+}\left( x,x^{\prime }\right) &=&i\int_{-\infty }^{+\infty
}dp_3dp_1\sum_{nr}\smallskip \ _{-}\psi _{p_3p_1nr}\left( x\right) \left[
g\left( \left. _{-}\right| ^{\_}\right) ^{-1}\right] ^{*}\smallskip \ ^{-}%
\bar{\psi}_{p_3p_1nr}\left( x^{\prime }\right) ,  \nonumber\\
S_{in}^{\mp }\left( x,x^{\prime }\right)& =&i\int_{-\infty }^{+\infty
}dp_3dp_1\sum_{nr}\smallskip \ _{\pm }\psi _{p_3p_1nr}\left( x\right) _{\pm }%
\bar{\psi}_{p_3p_1nr}\left( x^{\prime }\right) ,  \nonumber\\
S_{out}^{\mp }\left( x,x^{\prime }\right) &=&i\int_{-\infty }^{+\infty
}dp_3dp_1\sum_{nr}\smallskip \ ^{\pm }\psi _{p_3p_1nr}\left( x\right)
\smallskip \ ^{\pm }\bar{\psi}_{p_3p_1nr}\left( x^{\prime }\right) .
\label{w34}
\end{eqnarray}
 $\sum_{nr}$ means the summation over all discrete quantum
numbers $n,r$ (and the integration over the continuous $p_2$ if
$H=0$).
 Using the relations between GF and between
the matrices $G\left( {}_\zeta |{}^{\zeta ^{\prime }}\right) $ one can
present the functions $S^{\mp },$ $S_{in}^{\mp }$ and $S_{out}^{\mp }$ as
follows 
\begin{eqnarray}
\ &&\pm S^{\mp }(x,x^{\prime })=S^c(x,x^{\prime })\pm \theta (\mp
(x_0-x_0^{\prime }))S(x,x^{\prime }) ,  \nonumber \\
\ &&\pm S_{in}^{\mp }(x,x^{\prime })=S_{in}^c(x,x^{\prime })\pm \theta (\mp
(x_0-x_0^{\prime }))S(x,x^{\prime }) ,  \nonumber \\
\ &&\pm S_{out}^{\mp }(x,x^{\prime })=S_{out}^c(x,x^{\prime })\pm \theta
(\mp (x_0-x_0^{\prime }))S(x,x^{\prime }) ,  \nonumber \\
\ &&S_{in}^c(x,x^{\prime })=S^c(x,x^{\prime })-S^a(x,x^{\prime }) ,
\;\;
S_{out}^c(x,x^{\prime })=S^c(x,x^{\prime })-S^p(x,x^{\prime }) ,
\nonumber  \\
\ &&S^a(x,x^{\prime })=-i\int_{-\infty }^{+\infty
}dp_3dp_1\sum_{nr}\smallskip \ _{-}{\psi }_{p_3p_1nr}(x)\,\left[
g(_{+}|^{-})g(_{-}|^{-})^{-1}\right] ^{\dagger }{_{+}\bar{\psi}}%
_{p_3p_1nr}(x^{\prime }) ,  \nonumber \\
\ &&S^p(x,x^{\prime })=i\int_{-\infty }^{+\infty
}dp_3dp_1\sum_{nr}\smallskip \ ^{+}{\psi }_{p_3p_1nr}(x)\,\left[
g(_{+}|^{+})^{-1}g(_{+}|^{-})\right] {^{-}\bar{\psi}}_{p_3p_1nr}(x^{\prime
})\quad .  \label{w41}
\end{eqnarray}

Let us consider the functions
 $S^{\pm}$ 
and $S^{a,p}$. 
The coefficients (\ref{a15}) do not depend on $p_z$, thus, one can present
the functions $S^{\mp }$ and $S^{a,p}$ in the following convenient form 
\begin{equation}
S^{\mp ,a,p}(x,x^{\prime }) =\int_{-\infty }^{+\infty }\,dz\,\int_{-\infty
}^{+\infty }\,\frac{dp_3}{2\pi }\,e^{-ip_3y^3}S_Q^{\mp ,a,p}\;,\quad y_\mu
=x_\mu -x_\mu ^{\prime }\;,  \label{a24} 
\end{equation}
where
\begin{eqnarray}
S_Q^{-} &=&i\int_{-\infty }^{+\infty }dp_zdp_1\sum_{nr}{}\smallskip \
^{+}\psi _{p_3p_1nr}(\widetilde{\eta },x_{\perp },z,p_z)g\left( \left.
_{+}\right| ^{+}\right) ^{-1}\smallskip \ _{+}\bar{\psi}_{p_3p_1nr}\left( 
\widetilde{\eta ^{\prime }},x_{\perp }^{\prime },z^{\prime },p_z\right) , 
\nonumber \\
S_Q^{+} &=&i\int_{-\infty }^{+\infty }dp_zdp_1\sum_{nr}\smallskip \ _{-}\psi
_{p_3p_1nr}\left( \widetilde{\eta },x_{\perp },z,p_z\right) \left[ g\left(
\left. _{-}\right| ^{\_}\right) ^{-1}\right] ^{*}\smallskip \ ^{-}\bar{\psi}%
_{p_3p_1nr}\left( \widetilde{\eta ^{\prime }},x_{\perp }^{\prime },z^{\prime
},p_z\right) ,  \nonumber \\
\ S_Q^a &=&-i\int_{-\infty }^{+\infty }dp_zdp_1\sum_{nr}\smallskip \ _{-}{%
\psi }_{p_3p_1nr}(\widetilde{\eta },x_{\perp },z,p_z)\,\left[
g(_{+}|^{-})g(_{-}|^{-})^{-1}\right] ^{\dagger }{_{+}\bar{\psi}}_{p_3p_1nr}(%
\widetilde{\eta ^{\prime }},x_{\perp }^{\prime },z^{\prime },p_z),  \nonumber
\\
S_Q^p &=&i\int_{-\infty }^{+\infty }dp_zdp_1\sum_{nr}\smallskip \ ^{+}{\psi }%
_{p_3p_1nr}(\widetilde{\eta },x_{\perp },z,p_z)\,\left[
g(_{+}|^{+})^{-1}g(_{+}|^{-})\right] {^{-}\bar{\psi}}_{p_3p_1nr}(\widetilde{%
\eta ^{\prime }},x_{\perp }^{\prime },z^{\prime },p_z),  \label{a26}
\end{eqnarray}
and
\begin{eqnarray}
&&_{\pm }\psi _{p_3p_1nr}\left( \widetilde{\eta },x_{\perp },z,p_z\right)
=\left( \gamma ^0i\partial _0+\gamma ^3\left( p_3-qEx^o\right) +\gamma
_{\bot }\left( i\partial -qA\right) +M\Omega \right)
 _{\pm }\phi _{p_3p_1nr}, \nonumber \\
&&_{\pm }\phi _{p_3p_1nr}
=\smallskip \ _{\pm }\phi _{p_3nr}(\widetilde{\eta },z,p_z)\,\phi
_{p_1nr}(x_{\perp })v_{+1,r}, \;
_{\pm }\phi _{p_3nr}(\widetilde{\eta },z,p_z)=\frac 1{\sqrt{2\pi }}%
\,e^{-ip_zz}\smallskip \ _{\pm }\phi _{p_3,n,+1,r}(x^0,p_z),  \label{H29}
\end{eqnarray}
and in the same form with $\left( \pm \right) $ indices above. Within
the same approximation one can rewrite the functions $%
S_Q^{\mp ,a,p}$ as follows:
\begin{equation}
S_Q^{\mp ,a,p}=\left( \gamma ^0i\partial _0+\gamma ^3\left( p_3-qEx^o\right)
+\gamma _{\bot }\left( i\partial -qA\right) +M\Omega \right) \bigtriangleup
_Q^{\mp ,a,p},  \label{H30}
\end{equation}
where the functions $\Delta _Q^{\mp ,a,p}$ obey the equation 
\[
\left[
\left( i\frac \partial {\partial \widetilde{\eta }}\right)
^2-\left( i\frac \partial {\partial z}-\rho \widetilde{\eta }\right) ^2+%
{\cal P}_{\perp }^2
-\frac{p_3^2\left( bM^2\right)} {\rho ^2} -i\rho \Xi
-\frac{q}{2}
F_{\mu \nu }^{\perp }\sigma ^{\mu \nu }\right] \Delta _Q^{\mp ,a,p}=0.
\]
They may
be easily expressed via the solutions (\ref{H29}) 
 according to the Eqs. (\ref{a26}). The functions $%
\Delta _Q^{\mp ,a,p}$ are just the GF
 of the
squared Dirac equation in electromagnetic background \cite{GGS,GGG}, where $%
\tilde{\eta}$ is the time, $z$ is the coordinate along the electric field,
the mass $m_Q^2=p_3^2(bM)^2/\rho ^2$, the potential of the electromagnetic
field is $A_z=\tilde{\eta}\rho /q$, and the spin term $qE\gamma ^0\gamma ^3$ is
changed to $\rho \Xi $. In what follows we are going to use
the following representations  \cite{GGG}:

\begin{eqnarray}
\ &&{\pm \Delta _Q^{\mp }}=\Delta _Q^c\pm \theta (\mp y_0)\Delta _Q\; ,
\nonumber \\[0.3cm]
\ &&\Delta _Q^c=\int_{\Gamma _c}\,f_Qds\; ,\;\;
\Delta _Q=\mbox{sgn}(y_0)\ \int_{\Gamma _c-\Gamma _2-\Gamma
_1}\,f_Qds\;,\nonumber
 \\[0.3cm]
\ &&\Delta _Q^a=\int_{\Gamma _a}\,f_Qds+\theta (z^{\prime }-z)\,\int_{\Gamma
_3+\Gamma _2-\Gamma _a}f_Qds\;,  \nonumber \\
\ &&\Delta _Q^p=\int_{\Gamma _a}\,f_Qds+\theta (z-z^{\prime })\,\int_{\Gamma
_3+\Gamma _2-\Gamma _a}f_Qds\;,  \label{a33a}
\end{eqnarray}
where $\theta (0)=1/2$, the contours of the integration are 
indicated on the Fig.1, and

\begin{figure}[h]
\begin{picture}(350,245)
\put(0,180){\vector(1,0){330}}
\put(325,185){Re $s$}
\put(100,0){\vector(0,1){240}} 
\put(107,233){Im $s$}
{\thicklines
\put(0,180){\vector(1,0){200}}
\put(0,110){\vector(1,0){200}}
\put(0,40){\vector(1,0){200}}
\put(0,180){\vector(-1,0){5}}
\put(0,40){\vector(-1,0){5}}
}
\put(140,185){$\Gamma_c$}
\put(140,115){$\Gamma_2$}
\put(140,45){$\Gamma_a$}
\put(40,185){$\Gamma_1$}
\put(40,45){$\Gamma_3$}
\put(220,110){$-i\frac{\pi}{2\rho}$}
\put(220,40){$-i\frac{\pi}{\rho}$}
\put(100,180){\circle*{5}}
\put(100,40){\circle*{5}}
\put(107,187){O}
\end{picture}
\caption[f1]{\label{f1}{Contours of integration 
$\Gamma_1,\Gamma_2,\Gamma_3,\Gamma_c,\Gamma_a$}}
\end{figure}
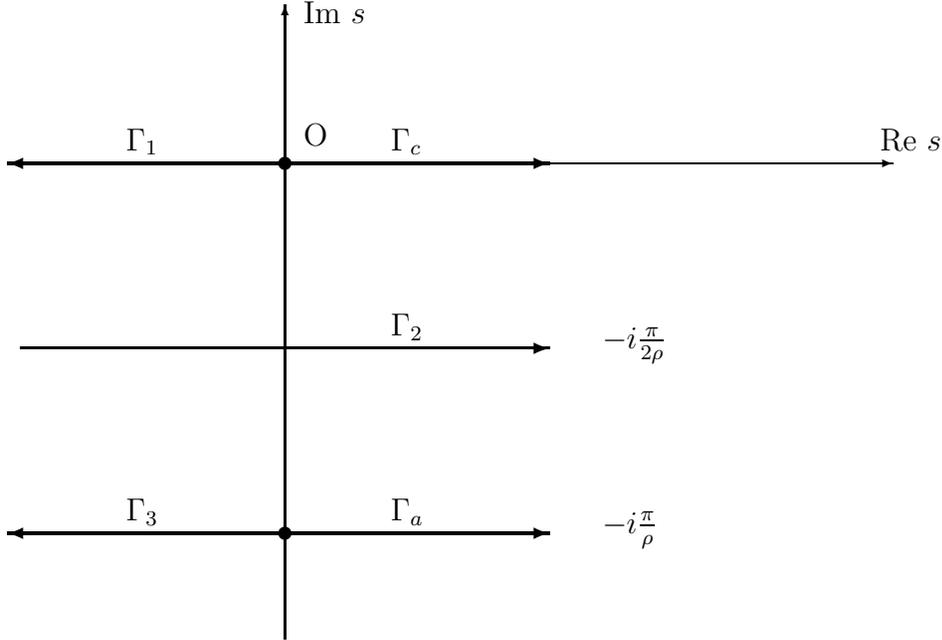

\begin{eqnarray}
&&f_Q =e^{s\rho \Xi }\exp \left( -\frac i2qF_{\mu
\nu }^{\perp }\sigma ^{\mu \nu }s\right)  f_Q^{(0)} ,  \label{H35}\\
&&f_Q^{(0)}=\exp 
\left\{-iq\int_{x^{\prime }}^x\,A_\mu ^{\perp }dx^\mu
\right\}f_Q^{\parallel }\left( %
z-z^{\prime }\right) f_{\perp } ,
\nonumber \\
&&f_{\perp }=(4\pi )^{-2}\,\frac{qH}{\sin (qHs)}\exp \left\{ -%
\frac i4y_{\perp }qF\coth (qFs)y_{\perp }\right\} \,,\;\;\;\;\;\;\; 
\nonumber \\
&&f_Q^{\parallel }\left(z\right) =\frac \rho {\sinh (\rho s)}\,\exp \left\{ -i\frac \rho 2(%
\widetilde{\eta }+\widetilde{\eta ^{\prime }})z-im_Q^2s
 +i\frac \rho 4\left[ z^2-(\widetilde{\eta }-\widetilde{\eta
^{\prime }})^2\right] \coth (\rho s)\right\}  .  \nonumber
\end{eqnarray}
One can calculate in  (\ref{a24}) 
all Gaussian integrals over 
$p_3$ and $z$.  Thus, one
gets 
\begin{equation}
S^{(\ldots )}\left( x,x^{\prime }\right) =\left( \gamma ^\mu {\cal P}_\mu
+M\Omega\right) \bigtriangleup ^{(\ldots )}\left( x,x^{\prime }\right) ,
\label{H37}
\end{equation}
\begin{eqnarray}
&&\Delta ^c(x,x^{\prime })=\int_{\Gamma _c}f(x,x^{\prime },s)ds ,\;\;
\Delta (x,x^{\prime })=
\mbox{sgn} (y_0)\int_\Gamma f(x,x^{\prime },s)ds
,  \nonumber \\[0.3cm]
&&\Delta ^a(x,x^{\prime })=-\Delta ^{(1)}(x,x^{\prime })-\Delta
^{(2)}(x,x^{\prime }) ,  \nonumber \\
&&\Delta ^p(x,x^{\prime })=-\Delta ^{(1)}(x,x^{\prime })+\Delta
^{(2)}(x,x^{\prime }) ,  \nonumber \\
&&\Delta ^{(1)}(x,x^{\prime })=-\frac 12\,\int_{\Gamma _3+\Gamma _2+\Gamma
_a}\,f(x,x^{\prime },s)ds,  \nonumber \\
&&\Delta ^{(2)}(x,x^{\prime })=\int_{\Gamma _3+\Gamma _2-\Gamma
_a}f_r(x,x^{\prime },s)ds ,  \label{a42a}
\end{eqnarray}
where
$f_r(x,x^{\prime },s)
=\pi^{-1/2}/2\;\gamma \left( 1/2,\alpha
\right) f(x,x^{\prime },s)$,  $\gamma \left( 1/2,\alpha \right) $ is
the incomplete gamma-function, and  
$\alpha =e^{-i\pi /2}\left(4s(bM)^2\omega \right)^{-1}
\left[ \left( x_0+{x^{\prime}}_0\right) s(bM)^2+qEy^3\right] ^2$.  
Here 
\begin{eqnarray}
&&f\left( x,x^{\prime },s\right) =
 \exp \left( \rho \Xi s-i\frac q2\sigma ^{\mu \nu
}F_{\mu \nu }^{\perp }s\right)
f^{(0)}\left( x,x^{\prime },s\right) ,
\label{a43} \\
&&f^{(0)}(x,x^{\prime },s)=\exp 
\left\{-iq\int_{x^{\prime }}^x\,A_\mu dx^\mu
\right\}
f_{\parallel }
 \,f_{\perp } ,  \nonumber \\
&&f_{\parallel } =\frac \rho {{\rm %
sinh}(\rho s)\omega ^{1/2}}\exp \left\{ i\frac{qE}2\left( x_0+{x^{\prime }}%
_0\right) y^3-i\frac \rho 4\left( x_0-{x^{\prime }}_0\right) ^2{\rm coth}%
(\rho s)\right.   \nonumber \\[0.3cm]
&&\left. -i(aM)^2s+i\frac \rho {4\omega }y_3^2\,{\rm coth}(\rho s)-\frac i{%
4\omega }\left[ (bM)^2s\left( x_0+{x^{\prime }}_0\right) ^2+2qEy^3\left( x_0+%
{x^{\prime }}_0\right) \right] \right\}  ,  \nonumber 
\end{eqnarray}
where $\omega =
s\,{\rm coth}(\rho s)(bM)^2/\rho
+(qE)^2/\rho ^2$. 
 One can see that 
\begin{eqnarray}  \label{a51}
&&i\frac d{ds}\,f(x,x^{\prime },s)=\left( M^2\Omega ^2-{\cal P}^2+\frac q2%
\sigma ^{\mu \nu }F_{\mu \nu }-ibM\gamma ^0\right) f(x,x^{\prime },s)\quad ,
\label{a52} \\[0.3cm]
&&\lim_{s\rightarrow +0}f(x,x^{\prime },s)=i\delta ^{(4)}(x-x^{\prime
})\quad .
\end{eqnarray}
Thus, $f(x,x^{\prime },s)$ is the Fock-Schwinger function \cite{Sch1,Fock}
of the QED-$\Omega $ theory, and $s$ is the Fock-Schwinger proper-time in
this case. The contour $\Gamma _c-\Gamma _2-\Gamma _1$ in representation
(\ref{a42a}) for $\Delta$
transformed into $\Gamma $ (see Fig.2)
after the integration over $p_D$ and $z$. 
The result 
 is consistent with the general expression for the commutation
function obtained in \cite{GG2}. The function $f^{(0)}\left( x,x^{\prime
},s\right) $ which has
appeared in (\ref{a43}) coincides with the Fock-Schwinger
function of the scalar case. Due to that we can widely use results 
presented in \cite{GGO}.

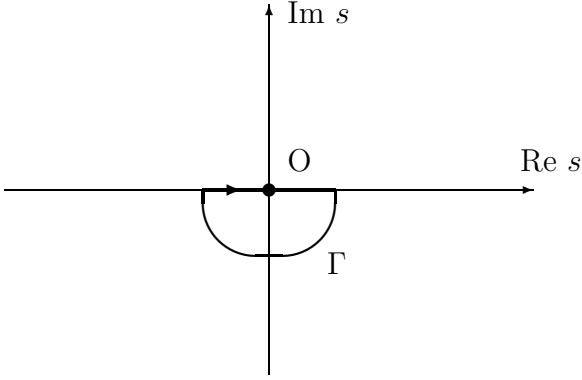
\begin{figure}[h]
\begin{picture}(210,150)
\put(0,70){\vector(1,0){200}}
\put(195,77){Re $s$}
\put(100,0){\vector(0,1){140}} 
\put(107,133){Im $s$}
{\thicklines
\put(100,70){\oval(50,50)[b]}
\put(122,38){$\Gamma$}
\put(85,70){\vector(1,0){5}}
\put(75,70){\line(1,0){50}}
}
%
\put(100,70){\circle*{5}}
\put(107,77){O}
\end{picture}
\caption[f2]{\label{f2}{Contour of integration $\Gamma$}}
\end{figure}

If $b\neq 0$, then the function $f(x,x^{\prime },s)$ has three singular
points on the complex region between contours $\Gamma _c-\Gamma _1$ and $%
\Gamma _a-\Gamma _3$, which are distributed at the imaginary axis: $\rho
s_0=0,$ $\rho s_1=-i\pi $ and $\rho s_2=-ic_2.$ The latter point is
connected with zero value of the function $\omega $. We get an equation for $%
c_2$ from the condition $\omega =0$, 
$
c_2\tan (c_2-\pi /2)-\left( qE/(bM)\right) ^2=0\;,  
$
where $\pi/2<c_2<\pi$. The position of this point depends on the ratio $%
qE/(bM)$, e.g. $c_2\rightarrow \pi$
as $bM/(qE)\rightarrow 0$  and $c_2\rightarrow \pi/2$
as 
$qE/(bM)\rightarrow 0$.  Notice,  in the case $%
E=0$ it is convenient to put $c_2=\pi/2+0$ since the contour $\Gamma_2$
is also passed above the singular point $s_2$. 
If $b=0$, then $\omega=1$, and the function $f(x,x^{\prime},s)$ has only two
singular points $s_0$ and $s_1$ on the above mentioned complex
region. In  this degenerate case we have known representation \cite{GGG}.

Similarly to Ref.
 \cite{GGO} one may demonstrate that the function $\Delta
^{(2)}$ from (\ref{a42a}) can  be presented via 
proper-time integrals with the Fock-Schwinger kernel $f(x,x^{\prime },s)$.
Then, using (\ref{a52}), one
can verify that the representations  (\ref{a42a}) for $\Delta^{(1)}$
 and $\Delta^{(2)}$
obey the
equation (\ref{H3b}). Thus, all the $\Delta $-functions considered
here, excluding those marked by the index ``c'', are solutions of the
equation (\ref{H3b}). The important difference between the functions $\Delta
^c$, $\Delta ^{(1)}$ and $\Delta$,
 $\Delta ^{(2)}$ is that the first ones are symmetric
under simultaneous change of sign in $x_0,\;x_0^{\prime
},\;x_3,\;x_3^{\prime }$ and the second ones change sign in this case.

Using the kernel (\ref{a43}) one can express
out-in effective action in the QED-$\Omega $ theory,

\begin{equation}
\Gamma _{out-in} =\frac 12\int dx\int_0^\infty s^{-1}\,f(x,x,s)\,ds.
\end{equation}


\section{Vacuum instability, mean values of current, and energy
  momentum tensor. Discussion}

Using the exact solutions and the GF constructed above, one may calculate
different physical effects having both local and global character. The
proper-time representation of the GF gives us a possibility to study
all mean values of current and energy momentum tensor in the
same manner.

All the information about the processes of particles creation, annihilation,
and scattering in an external field (without radiative corrections) can be
extracted from the matrices $G\left( {}_\zeta |{}^{\zeta ^{\prime }}\right) $
(\ref{we10}). These matrices define a canonical transformation between in and
out creation and annihilation operators in the generalized Furry
representation \cite{GG1977,FGS}, 
\begin{equation}
a^{\dagger }(out)=a^{\dagger }(in)G\left( {}_{+}|{}^{+}\right)
+b(in)G\left( {}_{-}|{}^{+}\right) , \;\;
b(out)=a^{\dagger }(in)G\left( {}_{+}|{}^{-}\right) +b(in)G\left(
{}_{-}|{}^{-}\right) .
\end{equation}
Here $a_l^{\dagger }(in)$, $b_l^{\dagger }(in)$, $a_l(in)$, $b_l(in)$ are
creation and annihilation operators of in-particles and antiparticles
respectively and $a_l^{\dagger }(out)$,$b_l^{\dagger }(out)$, $%
a_l(out),\;b_l(out)$ are ones of out-particles and antiparticles, $l$ are
possible quantum numbers (in our case it  $l=p_1,p_3,n,r$) . For
example, the mean differential
number of particles created (which are also equal to the
number of pairs created) by the external field from the in-vacuum $|0,in>$
is 
\begin{equation}
N_l=<0,in|a_l^{\dagger }(out)a_l(out)|0,in>=\left| g\left(
{}_{-}|{}^{+}\right) \right| ^2  \label{e11}
\end{equation}
(for a review of gravitational particles creation, see \cite{GMM,Par}). The
standard space coordinate volume regularization was used to get the latter
formula, so that $\delta (p_j-p_j^{\prime })\rightarrow \delta
_{p_j,p_j^{\prime }}$. The probability for a vacuum to remain a vacuum is 
\begin{equation}
P_v=|c_v|^2=\exp \left\{ \sum_l\ln \left( 1-N_l\right) \right\} \;.
\label{egs13}
\end{equation}

Similar to the electric field case \cite{GG1} we get
\begin{equation}
N_l==e^{-\pi \lambda },\;\;\mbox{if}\quad \sqrt{\rho }\,T\gg 1,\;\;\mbox{and}%
\;\sqrt{\rho }T\gg \lambda ,\;\mbox{and}\;\rho ^2T\gg |qEp_3|\;,  \label{ags14}
\end{equation}
where $\lambda $ is defined in (\ref{a12}). The latter conditions take place
with large enough time for action
of the electric-like field, 
$T=x_{out}^0-x_{in}^0$.
The result (\ref{ags14}) coincides with  one obtained in \cite{BO1}, and for 
$b=0$  it coincides with  one
obtained in \cite{GG1}.
Evidently the creation process is a coherent effect of both the
electromagnetic and gravitational fields. 
If the condition $p_3^2(bM)^2/\rho ^3<<1$ takes place (the gravitational
field is in a sense weaker than the electric one), the $p_3$
dependence on $N_l$
 is similar to the case $b=0$.
Thus , one can estimate that $\int dp_3=(qE)^{-1}\rho ^2T$ \cite{GG1}. Then
 the particle creation per unit of time may be calculated
similar to \cite
{SD}. In strong enough gravitational background the time dependence of the
effect is nonlinear  and needs to be studied specially.

To get the total number $N$ of particles created one has to sum over the
quantum numbers $l$. The sum over the momenta can be easily transformed into
an integral. Thus, if $b=0$ one gets result presented in \cite{GG1}. If $%
b\neq 0,$ the total number of pairs created per space coordinate volume has
the form 
\begin{equation}
\tilde{n}^{cr}=\frac{\sum_lN_l}{\int d{\bf x}}=\frac{\beta (1)}{4\pi ^2}%
\frac{\rho ^{3/2}}{bM}\exp \left\{ -\pi \frac{(aM)^2}\rho \right\} ,
\label{ags82}
\end{equation}
where $\beta (n)=qH\coth (n\pi qH/\rho ).$ The observable number density of
the created pairs in the asymptotic region $x_0=x_0^{out}\rightarrow \infty $
is given by the expression
$
{n}^{cr}=\tilde{n}^{cr}/\Omega ^3(x_0)\;.  
$
If the electromagnetic field is absent and $a=0$ , these results coincide
with ones in \cite{AS}.
In case $b\rightarrow 0$ the expression (\ref{ags82}) is growing infinitely.
In this case the particles are created in main by the electric field,
whereas the parameter $b$ plays a role of ``cut-off'' factor, which
eliminates creation of particles with extremely high momenta along the
electric field. 
It is seen from the expression (\ref{ags14}). Thus, the
limit $b\rightarrow 0$  corresponds to the case of the
electric field which acts for infinite time. Then the number of particles
created is proportional to the time of the field action. As was already
remarked above, in this case $\int dp_3=(qE)^{-1}\rho ^2T.$ Then, 
the parameter $b$ may be understood as
$(\sqrt{\rho })(TM)^{-1}$.
The vacuum-to-vacuum transition probability can be calculated, using formula
(\ref{egs13}). Thus, we get an analog of the well-known Schwinger formula \cite
{Sch1} in the case under consideration,
\begin{equation}
P_v=\exp \left\{ -\mu \tilde n^{cr}\int d{\bf x}\right\} ,\;\;\mu
=\sum_{n=0}^\infty \frac{\beta (n+1)}{(n+1)^{3/2}\beta (1)}\exp \left\{
-n\pi \frac{(aM)^2}\rho \right\} \;.  \label{add3}
\end{equation}

Now we are going to
 discuss vacuum matrix elements of current and
 of metric energy-momentum tensor (EMT) of spinor field.
  Making the conformal transformation the current in FRW universe
may be presented in the following  form 
\begin{equation}
\;J_\mu =\Omega ^{-2}(x^0)j_\mu ,\smallskip \smallskip \ j_\mu =\frac q2%
\left[ \bar{\psi}(x),\gamma _\mu \;\psi (x)\right] ,
\end{equation}
where $j_\mu $ is QED-$\Omega $ current  of the spinor
field $\psi (x)$. The latter 
 obeys the Dirac equation (\ref{H1}). The  EMT of
spinor field in FRW universe may be presented as 
\begin{eqnarray}
\  &&\tau _{\mu \nu }=\Omega ^{-2}(x^0)T_{\mu \nu },\smallskip \smallskip \
T_{\mu \nu }=\frac 12\left( T_{\mu \nu }^{can}+T_{\nu \mu }^{can}\right) ,\;,
\label{w84} \\
\  &&T_{\mu \nu }^{can}=\frac 14\left\{ \left[ \bar{\psi}(x),\gamma _\mu 
{\cal P}_\nu \;\psi (x)\right] +\left[ {\cal P}_\nu ^{*}\bar{\psi}(x),\gamma
_\mu \;\psi (x)\right] \right\} \;,  \nonumber
\end{eqnarray}
where $T_{\mu \nu }$ is EMT in QED-$\Omega .$ In the case of
 unstable vacuum
one needs to calculate three types of matrix elements
 \cite{FGS}, depending on the problem in question: 
\begin{eqnarray}
\  &<&j_\mu >^c=<0,out|j_\mu |0,in>c_v^{-1}\;,\smallskip \ \ <T_{\mu \nu
}>^c=<0,out|T_{\mu \nu }|0,in>c_v^{-1}\;,  \label{gsf83} \\
\  &<&j_\mu >^{in}=<0,in|j_\mu |0,in>\;,\smallskip \ \ <T_{\mu \nu
}>^{in}=<0,in|T_{\mu \nu }|0,in>\;,  \label{gsf85} \\
\  &<&j_\mu >^{out}=<0,out|j_\mu |0,out>\;,\smallskip \ <T_{\mu \nu
}>^{out}=<0,out|T_{\mu \nu }|0,out>\;,  \label{gsf87}
\end{eqnarray}
 Using GF, which were found
before, one can  present these matrix elements as follows: 
\begin{eqnarray}
\  &<&j_\mu >^c=iq\mbox{tr}\left\{ \gamma _\mu S^c(x,x)\right\} =iq\left. %
\mbox{tr}\left\{ \gamma _\mu \left( \gamma ^\kappa {\cal P}_\kappa +M\Omega
\right) \Delta ^c(x,x^{\prime })\right\} \right| _{x=x^{\prime }}\;,
\label{w3.88} \\
\  &<&T_{\mu \nu }>^c=i/4\left. \mbox{tr}\left\{ \left( \gamma _\mu \left( 
{\cal P}_\nu +{\cal P^{\prime }}_\nu ^{*}\right) +\gamma _\nu \left( {\cal P}%
_\mu +{\cal P^{\prime }}_\mu ^{*}\right) \right) S^c(x,x^{\prime })\right\}
\right| _{x=x^{\prime }}  \nonumber \\
\  &=&i\left. \mbox{tr}\left\{ B_{\mu \nu }\Delta ^c(x,x^{\prime })\right\}
\right| _{x=x^{\prime }}\;,  \label{w3.89} \\
\  &<&j_\mu >^{in}=<j_\mu >^c+<j_\mu >^{(1)}+<j_\mu >^{(2)}\;,\;\;
\label{w3.90a} \\
\  &<&j_\mu >^{out}=<j_\mu >^c+<j_\mu >^{(1)}-<j_\mu >^{(2)}\;,\;\;
\label{3.90b} \\
\  &<&T_{\mu \nu }>^{in}=<T_{\mu \nu }>^c+<T_{\mu \nu }>^{(1)}+<T_{\mu \nu
}>^{(2)}\;,\;\;\;  \label{w3.91a} \\
\  &<&T_{\mu \nu }>^{out}=<T_{\mu \nu }>^c+<T_{\mu \nu }>^{(1)}-<T_{\mu \nu
}>^{(2)}\;,\;  \label{w3.91b} \\
\  &<&j_\mu >^{(1,2)}=iq\left. \mbox{tr}\left\{ \gamma _\mu \left( \gamma
^\kappa {\cal P}_\kappa +M\Omega \right) \Delta ^{(1,2)}(x,x^{\prime
})\right\} \right| _{x=x^{\prime }}\;,  \label{w3.92} \\
\  &<&T_{\mu \nu }>^{(1,2)}=i\left. \mbox{tr}\left\{ B_{\mu \nu }\Delta
^{(1,2)}(x,x^{\prime })\right\} \right| _{x=x^{\prime }}\;,  \label{w3.93} \\
\  &&B_{\mu \nu }=1/4\left\{ \gamma _\mu \left( {\cal P}_\nu +{\cal %
P^{\prime }}_\nu ^{*}\right) +\gamma _\nu \left( {\cal P}_\mu +{\cal %
P^{\prime }}_\mu ^{*}\right) \right\} \left( \gamma ^\kappa {\cal P}_\kappa
+M\Omega \right) \;,  \nonumber
\end{eqnarray}
where $\,{\cal P^{\prime }}_\mu ^{*}=-i\frac \partial {\partial x^{\prime
}{}^\mu }-q\,A_\mu (x^{\prime }),$ the GF are given by Eqs. (%
 (\ref{a42a}),  
and the relation 
$\Delta ^c(x,x)=(1/2)\left[ \Delta^{-}(x,x)-\Delta^{+}(x,x)\right] $ is used.

To get convenient forms of $ <j_\mu >^{in}$ 
and $<T_{\mu \nu}>^{in}$ we may
 rewrite $\Delta ^{(1)}$ as follows: 
\begin{eqnarray}
&&\Delta ^{(1)}(x,x^{\prime }) =-\frac 12\Delta ^{\Gamma _2}(x,x^{\prime
})+\Delta ^{(3)}(x,x^{\prime }),  \;\;
\Delta ^{\Gamma _2}(x,x^{\prime }) =\int_{\Gamma _2}f(x,x^{\prime },s)ds, 
\nonumber \\
&&\Delta ^{(3)}(x,x^{\prime }) =-\frac 12\int_{\Gamma _3+\Gamma
_a}f(x,x^{\prime },s)ds\;.  \label{gsf17}
\end{eqnarray}
The contour $\Gamma _2$ does not pass through
the singular points of the function $%
f(x,x^{\prime },s)$, thus the contributions from $\Delta ^{\Gamma
_2}$ to EMT and to the current are finite. Displacing the
contour $\Gamma _2$ to the real axis we may present
the difference
 $\Delta ^c-\Delta ^{\Gamma _2}$
 as 
\begin{eqnarray}
&&\Delta ^c(x,x^{\prime })-\Delta ^{\Gamma _2}
(x,x^{\prime })=\Delta ^{\bar{c}%
}(x,x^{\prime })+\Delta ^\Gamma (x,x^{\prime }),  \nonumber\\
&&\Delta ^{\bar{c}}(x,x^{\prime })=\int_{-0}^{-\infty }
f(x,x^{\prime },s)ds, \;\;
\Delta ^\Gamma (x,x^{\prime })=\int_\Gamma f(x,x^{\prime },s)ds,
\label{emt100}
\end{eqnarray}
where $\Delta ^{\bar{c}}$
is related to
the anticausal GF, $S^{\bar{c}}(x,x^{\prime
})=(\gamma {\cal P}+M\Omega )\Delta ^{\bar{c}}(x,x^{\prime })$, and the
integral  $\Delta ^\Gamma$ 
can be expressed by means of the $\Delta $-function
related to the commutation function, $%
S(x,x^{\prime })=(\gamma {\cal P}+M\Omega )\Delta (x,x^{\prime })$ \cite
{GG2}: 
$
\Delta ^\Gamma (x,x^{\prime })=\mbox{sgn}(x_0-x_0^{\prime })\Delta
(x,x^{\prime }). 
$
Taking into account the above definition the space-like part
of the limit $x-x^{\prime }\rightarrow 0$ in the EMT expressions have to be
treated either as the limit $x_0-x_0^{\prime }=+0$ or as $%
x_0-x_0^{\prime }=-0$, respectively. Then, in according to the initial
condition for the commutation function we get 
$
\left. \Delta (x,x^{\prime })\right| _{x_0=x_0^{\prime }}=0,\smallskip \
\smallskip \ \left. \partial_0 \Delta (x,x^{\prime
})\right| _{x_0=x_0^{\prime }}=\delta ({\bf x-x}^{\prime }).
$
Thus,  the contributions from $\Delta $ into $<T_{\mu \nu
}>^{in}$ and $<j_\mu >^{in}$ are zero. (All contributions from
$\Delta$, which might  appear as a result of a change of the limit
definition, are background independent. They may be eliminated by 
a renormalization.) It follows
from (\ref{a43}) that 
\begin{equation}
f(x,x^{\prime },-s^{*})=\gamma ^0f(x^{\prime },x,s)^{\dagger }\gamma ^0.
\label{gsf24}
\end{equation}
Changing $s\rightarrow -s$ one can represent integral
$\Delta ^{\bar{c}}$ from
(\ref{emt100}) in the form 
\begin{equation}
\Delta ^{\bar{c}}(x,x^{\prime })=-\int_0^\infty \gamma ^0f(x^{\prime
},x,s)^{\dagger }\gamma ^0ds,  \label{emt17b}
\end{equation}
to get

\begin{eqnarray}
&<&j_\mu >^{in}=%
\mathop{\rm Re}%
<j_\mu >^c+<j_\mu >^{(2)}+<j_\mu >^{(3)}\;,  \label{gsf18} \\
&<&T_{\mu \nu }>^{in}=%
\mathop{\rm Re}%
<T_{\mu \nu }>^c+<T_{\mu \nu }>^{(2)}+<T_{\mu \nu }>^{(3)}\;,  \label{gsf19}
\end{eqnarray}
where the terms $<j_\mu >^{(3)}$ and $<T_{\mu \nu }>^{(3)}$ represent the
contributions from $\Delta ^{(3)}$: 
\begin{eqnarray}
&<&j_\mu >^{(3)}=iq\left. \mbox{tr}\left\{ \gamma _\mu \left( \gamma ^\kappa 
{\cal P}_\kappa +M\Omega \right) \Delta ^{(3)}(x,x^{\prime })\right\}
\right| _{x=x^{\prime }}\;,  \label{gsf20} \\
&<&T_{\mu \nu }>^{(3)}=i\left. \mbox{tr}\left\{ B_{\mu \nu }\Delta
^{(3)}(x,x^{\prime })\right\} \right| _{x=x^{\prime }}\;.  \label{gsff21}
\end{eqnarray}
Using the Eq. (\ref{gsf24}) one can verify that each of the terms $<j_\mu
>^{(2)}, $ $<j_\mu >^{(3)},$ $<T_{\mu \nu }>^{(2)}$ and $<T_{\mu \nu
}>^{(3)} $ is real, as it follows from their initial definitions.

One can get similar expressions for the
out-out-matrix elements of the current and EMT: 
\begin{eqnarray}
&<&j_\mu >^{out}=%
\mathop{\rm Re}%
<j_\mu >^c-<j_\mu >^{(2)}+<j_\mu >^{(3)}\;,  \label{gsf22} \\
&<&T_{\mu \nu }>^{out}=%
\mathop{\rm Re}%
<T_{\mu \nu }>^c-<T_{\mu \nu }>^{(2)}+<T_{\mu \nu }>^{(3)}\;.  \label{gsf23}
\end{eqnarray}

All
nondiagonal matrix elements of EMT are zero: 
\begin{equation}
<T_{\mu \nu }>^c=<T_{\mu \nu }>^{(2)}=<T_{\mu \nu }>^{(3)}=0,\;\mu \neq \nu .
\end{equation}
Using the formulas 
\begin{eqnarray*}
\exp \left( \rho \Xi s\right)  &=&\cosh \left( \rho s\right) +\Xi \sinh
\left( \rho s\right) ,  \nonumber \\
\exp \left( -i\frac q2\sigma ^{\mu \nu }F_{\mu \nu }^{\perp }s\right) 
&=&\cos \left( qHs\right) +\gamma ^2\gamma ^1\sin \left( qHs\right), 
\end{eqnarray*}
one may calculate the traces for the current
and EMT: 
\begin{eqnarray*}
&&    \tau (s)= \mbox{tr}Y =4\cosh \left( \rho s\right) \cos
\left( qHs\right) , \;\;
Y=\left\{ \exp 
\left( \rho \Xi s-i\frac q2\sigma ^{\mu \nu
}F_{\mu \nu }^{\perp }s\right) \right\},    \\
&&\mbox{tr}\left\{ \Xi Y \right\} =\tanh (\rho s)\tau (s)\;, \;\;
\mbox{tr}\left\{ \gamma ^0\gamma ^3 Y 
\right\} =\frac{qE}\rho \tanh (\rho
s)\tau (s)\;, \\
&&\mbox{tr}\left\{ \gamma ^0 Y
 \right\} =-\frac{bM}\rho \tanh (\rho s)\tau
(s)\;, \;\;
\mbox{tr}\left\{ \gamma ^2\gamma ^1 Y
 \right\} =-\tan (qHs)\tau
(s)\;,
\end{eqnarray*}
and other traces are zero.
Then, the current matrix elements can be found in the forms 
\begin{eqnarray}
&<&j_\mu >^c=\int_{\Gamma _c}\left. \alpha _\mu (s)\tau
(s)f^{(0)}(x,x^{\prime },s)ds\right| _{x=x^{\prime }}\;,  \label{gsf1} \\
&<&j_\mu >^{(3)}=-\frac 12\int_{\Gamma _3+\Gamma _a}\left. \alpha _\mu
(s)\tau (s)f^{(0)}(x,x^{\prime },s)ds\right| _{x=x^{\prime }}\;,  \label{gsf2}
\\
&<&j_\mu >^{(2)}=\int_{\Gamma _3+\Gamma _2-\Gamma _a}\left. \alpha _\mu
(s)\tau (s)f_r^{(0)}(x,x^{\prime },s)ds\right| _{x=x^{\prime }}\;,
\label{gsf3} \\
&&\alpha _\mu (s)=iq\delta _\mu ^3\left[ {\cal P}_3+{\cal P}_0\frac{qE}\rho
\tanh (\rho s)\right] \;,  \nonumber
\end{eqnarray}
where the only $x^3$ components (along the electric field) of the currents
differ from zero and vanish in the absence of the electric field, as it
presented to be. The nonzero components of $<T_{\mu \nu }>^c$ are 
\begin{eqnarray}
\  &<&T_{\mu \nu }>^c=\int_{\Gamma _c}t_\mu (s)\tau
(s)f^{(0)}(x,x,s)ds\;,\;\;\mbox{if}\;\mu =\nu \;,  \label{gsf4} \\
\  &&t_0(s)=-\frac \rho {\sinh (2\rho s)},\;\;t_2(s)=t_1(s)=\frac{qH}{\sin
(2qHs)}\;,  \nonumber \\
&&t_3(s)=\frac \rho {2\omega }\coth (\rho s)-\frac{(qE)^2}{2\rho }\tanh
(\rho s)+i\left( \frac{qE}2(x_0+x_0^{\prime })(1-\omega ^{-1})\right) ^2, 
\nonumber
\end{eqnarray}
and nonzero components of $<T_{\mu \nu }>^{(2)}$ and $<T_{\mu \nu
}>^{(3)}$ can be written as follows: 
\begin{eqnarray}
&<&T_{\mu \nu }>^{(3)}=-\frac 12\int_{\Gamma _3+\Gamma _a}l_\mu (s)\tau
(s)f^{(0)}(x,x,s)ds\;,\;\;\mbox{if}\;\mu =\nu \;,  \label{gsf5} \\
&<&T_{\mu \nu }>^{(2)}=\int_{\Gamma _3+\Gamma _2-\Gamma _a}\left. l_\mu
(s)\tau (s)f_r^{(0)}(x,x^{\prime },s)ds\right| _{x=x^{\prime }}\;,\;\;%
\mbox{if}\;\mu =\nu \;,  \label{gsf6} \\
&&l_1(s)=t_1(s),\;\;l_2(s)=t_2(s),\;l_3(s)=i\left( {\cal P}_3\right) ^2-%
\frac{(qE)^2}{2\rho }\tanh (\rho s)\;,  \nonumber \\
&&l_0(s)=l_1(s)+l_2(s)+l_3(s)+i\left[ 2\left( \omega ^{-1}b^2M^2sx_0\right)
^2-i\omega ^{-1}b^2M^2s+M^2\Omega ^2\right] -\frac \rho 2\tanh (\rho s)\;.\;
\nonumber
\end{eqnarray}
Here, $l_0(s)$ is obtained from $t_0(s)$, using Eq. (\ref{a52}) and taking
into account that the $\Delta ^{(2)}$ function obeys the equation (\ref{H3b}).

The components $<j_\mu >^c$ and $<T_{\mu \nu }>^c$ are expressed by
the causal
GF in the well-known Schwinger's form. They are sources of information
about 
local features of the theory. For the first time it is obtained
exactly with respect to the given external background. 
The  components $<j_\mu >^{(2,3)}$ and $%
<T_{\mu \nu }>^{(2,3)}$ are only related to global features of the theory
and characterize
 the vacuum instability. 
 Such  expressions cannot be calculated in the
frame of the perturbation theory with respect to the external background or
in the frame of the WKB method. 
Then, such terms has never been studied.
 If the parameter $b\neq 0$, the
only expression (\ref{gsf4}) for $<T_{\mu \nu }>^c$ has to be
regularized and then
renormalized. The expression (\ref{gsf1}) for $<j_\mu >^c$ is finite after the
regularization lifting. The terms $<j_\mu >^{(2,3)}$ and $<T_{\mu \nu
}>^{(2,3)}$ are also finite. Thus, as
one can see from Eqs. (\ref{gsf19}) and (%
\ref{gsf23}), the divergence in $<T_{\mu \nu }>^{in}$ and $<T_{\mu \nu }>^{out}
$  is the same as in $<T_{\mu \nu }>^c.$ That is consistent with
the fact that the ultraviolet divergences have a local nature and result (as
in the theory without an external field) from the leading local terms at $%
s\rightarrow +0$. Nevertheless,  as $b\rightarrow
0$ ( flat space limit)
the terms $<j_\mu >^{(2,3)}$ and $<T_{\mu \nu }>^{(2,3)}$ have electric
field dependent divergences defined by the dimensionless parameter $%
\beta =bM/qE.$ The nature of that is similar to
the  divergence of $\tilde{n}^{cr}$
for $b\rightarrow 0$. In the case $b=0$ the density  of 
excited vacuum states 
 is proportional to the time $T$.
 After a special regularization with respect to
time $T$ \cite{GG1} all of the terms become finite. 

Now we are going to
 discuss current, energy density and pressure of the
created particles. Let us introduce normalized values of current and
EMT, which may be easily connected to observable values in some appropriate
asymptotic region $x_0=x_0^{as}$,
\begin{equation}
j_\mu ^{cr}=\tilde{j}_\mu ^{cr}/\Omega ^3(x_0)\;,  \;\;
T_{\mu \nu }^{cr}=\tilde{T}_{\mu \nu }^{cr}/\Omega ^3(x_0)\;,  \label{gsf96}
\end{equation}
where according to the definitions (\ref{gsf85}) and (\ref{gsf87})
the corresponding densities of particles created per space-coordinates
volume are 
\begin{eqnarray}
&&\tilde{j}_\mu ^{cr}=\frac{\int d{\bf x}\left( <j_\mu >^{in}-<j_\mu
>^{out}\right) }{\int d{\bf x}}\;,\;\;x_0=x_0^{as},  \label{gsf97} \\
&&\tilde{T}_{\mu \nu }^{cr}=\frac{\int d{\bf x}\left( <T_{\mu \nu
}>^{in}-<T_{\mu \nu }>^{out}\right) }{\int d{\bf x}}\;,\;\;x_0=x_0^{as}\,.
\label{gsf98}
\end{eqnarray}
 Then, using
representations (\ref{gsf18}), (\ref{gsf19}) and (\ref{gsf22}), (\ref{gsf23})
one gets from (\ref{gsf97}) and (\ref{gsf98}), 
\begin{equation}
\tilde{j}_\mu ^{cr}=2<j_\mu >^{(2)}\;,\;\;\;
\tilde{T}_{\mu \nu }^{cr}=2<T_{\mu \nu }>^{(2)}\;,\;\;x_0=x_0^{as}.
\label{gsf100}
\end{equation}
The component $T_0^{cr\smallskip \ 0}=\Omega ^{-4}\tilde{T}_{00}^{cr}$ is
the energy density and $-T_\mu ^{cr\smallskip \ \nu }=\Omega ^{-4}\tilde{T}%
_{\mu \nu }^{cr}$ for $\mu =\nu =1,2,3$ are components of pressure which
are measured by the cosmic observer relative to the measured volume. In
contrast to that, measured energy and pressure taken per coordinate volume
are given by 
$
\check{T}_0^{cr\smallskip \ 0}=\Omega ^{-1}\tilde{T}_{00}^{cr},\smallskip \
\smallskip \ -\check{T}_\mu ^{cr\smallskip \ \nu }=\Omega ^{-1}\tilde{T}%
_{\mu \nu }^{cr},\smallskip \ \mu =\nu =1,2,3.  
$

We are going to analyze contributions to the quantities 
(\ref{gsf100}) at $x_0=x_0^{as}$. The leading asymptotics in $<j_\mu >^{(2)}$
and $<T_{\mu \nu }>^{(2)}$ at $x_0>>\sqrt{\rho }/(bM)$ are determinated by
the expressions
\begin{equation}
j_\mu >^{(2)}=\alpha _\mu (s_1)\left. \mbox{tr}\left\{ \Delta
^{(2)}(x,x^{\prime })\right\} \right| _{x=x^{\prime }}\;, \;\;
T_{\mu \mu }>^{(2)}=l_\mu (s_1)\left. \mbox{tr}\left\{ \Delta
^{(2)}(x,x^{\prime })\right\} \right| _{x=x^{\prime }}\;.
\end{equation}
An asymptotic expression for $\mbox{tr} \Delta
^{(2)} $ with $x\rightarrow x^{\prime }$  can 
be found
using the method described in the App. A of Ref.\cite{GGO} for the
scalar case. Then, one gets 
\[
\mbox{tr} \Delta ^{(2)}  =\frac{-i\tilde{n}%
^{cr}}{\rho (x_0+x_0^{\prime })} 
\exp \left\{ iq\Lambda +\frac{iqE}{2}(x_0+x_0^{\prime })y^3-\frac{\rho
^3(y_3)^2}{4\pi (bM)^2}+y_{\perp }
\frac{qF}{4}\cot \left( \pi qF/\rho \right)
y_{\perp }\right\}  ,  
\]
where $\tilde{n}^{cr}$ is defined in (\ref{ags82}). This expression is also
valid when $x_0<0,$ since the function $\Delta
^{(2)}$ is odd in $x_0$ as $x=x^{\prime }$ (see Sec.III ).
 Thus, one can see  when 
$x_0>>\sqrt{\rho }/(bM)$ the leading terms in $%
<j_\mu >^{(2)}$ and $<T_{\mu \nu }>^{(2)}$ are 

\begin{eqnarray}
&<&j_\mu >^{(2)}=-q^2E\rho ^{-1}\tilde{n}^{cr}\delta _\mu ^3\smallskip \ ;
\nonumber\\
&<&T_{00}>^{(2)}=\left[ \rho ^2x_0^2+a^2M^2+2qH\sinh ^{-1}\left( 2\pi
qH/\rho \right) \right] \frac{\tilde{n}^{cr}}{\rho x_0}\;,  \nonumber \\
&<&T_{11}>^{(2)}=<T_{22}>^{(2)}=qH\sinh ^{-1}\left( 2\pi qH/\rho \right) 
\frac{\tilde{n}^{cr}}{\rho x_0}\;,  \nonumber \\
&<&T_{33}>^{(2)}=\left[ (qEx_0)^2+\frac{\rho ^3}{2\pi (bM)^2}\right] \frac{%
\tilde{n}^{cr}}{\rho x_0}\;.  \label{gsf102}
\end{eqnarray}
Comparing our results with the scalar
theory case
\cite{GGO}, when $x_0\rightarrow \infty $ 
we may see the difference  
only in the 
factor $\tilde{n}^{cr}$ which enters
in the leading terms of $<j_\mu >^{(2)},$ $%
<T_{00}>^{(2)}$ and $<T_{33}>^{(2)}.$

Doubling the expressions  (\ref{gsf102}) according to
the Eqs. 
 (\ref{gsf100}), one gets the mean densities for current and
EMT of particles created. It turns out that these quantities are
proportional to the density of total number of particles and antiparticles
created ($2\tilde{n}^{cr}$) for the infinite time and do not change their
structure with increasing of $x_0$. The latter means that one can consider
all the expressions obtained at any fixed $x_0$ as asymptotic forms if $%
x_0\gg \sqrt{\rho }/bM$. In a strong background  $a^2M^2/\rho \leq 1$
and, therefore, such a time has to obey the condition $x_0>>a/b$. Thus,
in the strong background our asymptotic conformal time $x_0$, which is large
enough in quantum sense explained, corresponds to the large cosmological
time $t$.

Note, that one can neglect the second term in the brackets of the expression
(\ref{gsf102}) for $<T_{33}>^{(2)}$ at $bM/(qE)\leq 1$. Also one can neglect
both the term $a^2M^2$ in the brackets of the expression (\ref{gsf102}) for $%
<T_{00}>^{(2)}$ in case of strong  background $a^2M^2/\rho \leq 1$
and third term in the same expression if the magnetic field is not
strong enough, $qH/\rho \leq 1.$ The current density 
 $\tilde{j}_\mu ^{cr}=2<j_\mu >^{(2)}$ does not
depend on the asymptotic time. As $b\rightarrow 0$ this expression coincides
with the
one for flat space, $<\tilde{j}_\mu >^{cr}=-2q\tilde{n}^{cr}\delta _\mu
^3.$ The pressure component along the electric field direction $\tilde{T}%
_{33}^{cr}=2<T_{33}>^{(2)}$ is growing with time upon the action of the
field. However, if $qE/(bM)<<1$ then the asymptotic condition for $x_0$ is
consistent with the fact that the term $(qEx_0)^2$ in the expressions $%
<T_{33}>^{(2)}$ from (\ref{gsf102}) will not be dominant before large enough
time instant $x_0$. In this case one can neglect the contribution which
depends on the electric field, if the field is switched off at a fixed
time.

The components of the pressure in the directions, which are perpendicular to
the electric field, $\tilde{T}_{11}^{cr}=2<T_{11}>^{(2)}$ and $\tilde{T}%
_{22}^{cr}=2<T_{22}>^{(2)}$, decrease when the magnetic field increases. If
an electromagnetic field is absent, the pressure is isotropic according to
the symmetry of the space-time and to
the corresponding symmetry of the vacuum.
For $x^0\rightarrow \infty $ the remaining terms of the measured energy and
pressure of the created particles taken per coordinate volume are 
\[
\check{T}_0^{cr\smallskip \ 0}=\frac \rho b2\tilde{n}^{cr},\smallskip \ -%
\check{T}_3^{cr\smallskip \ 3}=\frac{(qE)^2}{\rho b}2\tilde{n}^{cr},
\]
and, if an electric field is absent the only remaining term is $\check{T}%
_0^{cr\smallskip \ 0}=2M\tilde{n}^{cr}.$ The last represents the total rest
mass per coordinate volume and coincides with the result of Ref.\cite{SD}.

Another problem is to study a back-reaction of particles created on
the electromagnetic field and metrics, or to be more correct, it is better
to say about back-reaction effects produced by both of particles created
from a vacuum and polarization of an unstable vacuum. 
To this end one needs to
use the expressions $<j_\mu >^{in}$ and $<T_{\mu \nu }>^{in}$ for all times $%
x_0$. The explicit proper-time expressions
found above give us promising tool for
accurate analysis of the total back-reaction related to the
 vacuum instability. In this approach  we do not need to select
phenomenologically
parts come from real particles and from a vacuum polarization, that
has been usually done in literature (see, for example,. review in
\cite{GMM}).
 Of course, to
find self-consistent solution taking into account the back-reaction 
one needs  numerical estimations (compare with pure
electromagnetic case, \cite{KES}). Keeping in mind  such an application,
 one needs to get preliminary information
 about the behavior of the expressions (%
\ref{gsf18}) and (\ref{gsf19}) in time and to 
select the leading components. To this end
 let us estimate the above mentioned expressions
 at characteristic large time, 
$x_0^2>>\rho /(bM)^2,$ and at characteristic small time,
 $x_0^2<<\rho /(bM)^2.$

Neglecting  divergent terms in $<T_{\mu \nu }>^c$ according to the
standard renormalization procedure, one can reduce Eq. (\ref{gsf19})
to the following finite form 
\begin{equation}
<T_{\mu \nu }>_{fin}^{in}=%
\mathop{\rm Re}%
<T_{\mu \nu }>_{fin}^c+<T_{\mu \nu }>^{(2)}+<T_{\mu \nu }>^{(3)}\;.
\label{gsf11}
\end{equation}
Here $<T_{\mu \nu }>_{fin}^c$ is a finite part of $<T_{\mu \nu
}>^c$. An estimation of  $<T_{\mu
\nu }>_{fin}^c$ can be made only after renormalization. Since  we
are here interested in to reveal  global features of
the theory, details of 
the renormalization problem together
with others, related to the renormalization, will be considered
 in the next paper. To
select contributions related to the vacuum instability, we note that the
functions $\Delta ^c$ (\ref{a42a}) and $\Delta
^{(3)}$ (\ref{gsf17}) with $x=x^{\prime }$ are even  in 
$x_0$. Thus, the functions $<T_{\mu \nu }>^c$ and $<T_{\mu \nu }>^{(3)}$ are
also even  and do not vanish when $x_0\rightarrow 0$, whereas the
functions $<j_\mu >^c$ and $<j_\mu >^{(3)}$ are odd  and vanish in the
limit. Moreover,  $<j_\mu >^c=<j_\mu >^{(3)}=0$ for all $x_0$ with $b=0$%
. The proper-time integral $\Delta ^{(2)}$ (\ref{a42a}) is 
odd in $x_0$ with $x=x^{\prime }$ and vanishes when $x_0\rightarrow 0$.
Thus, the expression $<T_{\mu \nu }>^{(2)}$ is also  odd in $x_0$
and vanishes in this limit. The term $<j_\mu >^{(2)}$ is not zero if
 $E\neq 0$. It is even in 
$x_0$ and  differs from zero when $x_0\rightarrow 0$.

The asymptotic behavior of $<j_\mu >^{(3)}$ and $<T_{\mu \nu }>^{(3)}$ is
defined by the asymptotic expression for $\Delta ^{(3)}.$
Using the method  presented in Appendix A of Ref. \cite{GGO}
one can verify that asymptotically 
\begin{equation}
\Delta ^{(3)}(x,x^{\prime })=\mbox{sign}(x_0)\Delta ^{(2)}(x,x^{\prime
})\,\quad ,  \label{Aa10}
\end{equation}
and therefore, 
\begin{equation}
<j_\mu >^{(3)}=\mbox{sign}(x_0)<j_\mu >^{(2)}\,,\;\;\;<T_{\mu \nu }>^{(3)}=%
\mbox{sign}(x_0)<T_{\mu \nu }>^{(2)}\;.  \label{gsf103}
\end{equation}
The expression (\ref{gsf1}) does not need to be renormalized, thus, one can
easily verify that the relation $<j_3>^c\sim x_0^{-1}\rightarrow 0$ holds
asymptotically. Then, the asymptotics of $<j_\mu >^{in}$ is determined by
the asymptotic behavior of $<j_\mu >^{(2)}$ and $<j_\mu >^{(3)}.$ 
If $x_0$ with the large modulus is negative, $<j_\mu >^{in}=%
\mathop{\rm Re}%
<j_\mu >^c$ and $<T_{\mu \nu }>_{fin}^{in}=%
\mathop{\rm Re}%
<T_{\mu \nu }>_{fin}^c$. 
Thus, in this regime
the vacuum instability in the
 time-dependent background does not affect the expressions. 
If $\beta =bM/(qE)<<1,$ the expression $%
\mathop{\rm Re}%
<T_{\mu \nu }>_{fin}^c$ remains finite with $b\rightarrow 0$,
 while the terms $%
<T_{\mu \nu }>^{(2)}$ and $<T_{\mu \nu }>^{(3)}$ diverge in the limit
since the global effect: the total density of excited vacuum
states in an electric field grows infinitely and its mean momentum along the
electric field  increases. In this case 
 the asymptotics of $<T_{\mu \nu }>_{fin}^{in}$ is determined by
asymptotic behavior of $<T_{\mu \nu }>^{(2)}$ and $<T_{\mu \nu }>^{(3)},$
i.e., by the asymptotic global properties of the theory.

Let us estimate $<j_\mu >^{in}$ and $<T_{\mu \nu }>_{fin}^{in}$ for 
small $x_0$. The terms $<j_\mu >^c$ and $<j_\mu >^{(3)}$ are odd in
 $x_0$. They vanish at $x_0\rightarrow 0.$ That is why the leading term in
(\ref{gsf18}) is $<j_\mu >^{(2)}.$ The leading contribution from the
 latter can
be represent as 
\[
<j_\mu >^{(2)}=\alpha _\mu (s_2)\left. \mbox{tr}\left\{ \Delta
^{(2)}(x,x^{\prime })\right\} \right| _{x=x^{\prime }}\;.
\]
Then, using the expressions (\ref{App4}) from the Appendix A, one gets 
\begin{equation}
<j_\mu >^{(2)}=qn^{(2)}\left[ 1+\left( c_2(bM)^2/(\rho qE)\right) ^2\right]
\delta _\mu ^3.  \label{gsf108}
\end{equation}
The expression (\ref{gsf108}) differs
essentially  from the
asymptotic one (\ref{gsf102}). That remains true in the quasi-flat
metrics ( $\beta =bM/(qE)<<1$), when  the vacuum instability results
only from
the electric field.  In the case of the
small time limit the expression
$<j_\mu >^{(2)}=q\tilde{n}^{cr}\delta _\mu ^3$ differs
by a sign from the asymptotics (\ref{gsf102}).  
The factor $\tilde{n}^{cr}$ is known from the asymptotics. That is
natural since  value of 
the time instant $x_0$ is large enough ($%
x_0+T/2>>(qE)^{-1/2}$ and $x_0+T/2>>(aM)^2(qE)^{-3/2}$ according to (\ref
{ags14}) ) with respect to
the initial time instant $(x_0^{in}=-T/2)$. Then the field action time
is large enough to obey the stabilization condition and the  density
of exited states has already 
 the asymptotic form. However,  such $x_0$
is still less than the asymptotic time for any  $\beta $.
Hence, at the time instant $x_0$ the vacuum  differs essentially from 
$|0,out>$ vacuum. Technically that means that at the small 
time,  $x_0<<\sqrt{qE}/(bM)$,
one cannot
obtain the normal form of operators
by means of the $\Delta ^{out}$-function and,
consequently, to use only the term $%
<j_\mu >^{(2)}$ for calculation of
 the mean current of quasiparticles, which are 
 created up to that time.
 The small time expressions one can use only to calculate the
back-reaction. Thus,  in the external background under consideration
a small time
back-reaction to the mean electromagnetic field 
is not a
screening (as it may be expected by analogy with a back-reaction of real
particles created) but an induction. Since the small time 
considered is related to
early stages of the Universe,  such  an observation may be 
important.

Since the function $<T_{\mu \nu }>^{(2)}$ is odd in $x_0$ and
vanishes as $x_0\rightarrow 0,$
the leading contribution to (\ref{gsf11}) is determined by $%
\mathop{\rm Re}%
<T_{\mu \nu }>_{fin}^c$ and $<T_{\mu \nu }>^{(3)}.$ However, if $\beta
=bM/(qE)<<1$, 
the situation is
more simple in the domain where $\beta ^{-2}>>qEx_0^2$, 
and  the leading contributions come from $<T_{\mu
\nu }>^{(3)}$ only. That is related to
the same  global effect,
 mentioned above in the asymptotic case, since the expression $%
\mathop{\rm Re}%
<T_{\mu \nu }>_{fin}^c$ remains finite with $b\rightarrow 0$,
 while the term $%
<T_{\mu \nu }>^{(3)}$ diverges in the limit. The total  density
of excited vacuum states  grows infinitely and the mean
momentum of the states
along the electric field  increases, as well. Taking
into account that
the leading contributions in $<T_{\mu \nu }>^{(3)}$ result from the
integral over a neighborhood of the singular point $s_1$, one gets 
\begin{eqnarray}
&<&T_{00}>^{(3)}=<T_{33}>^{(3)}=\frac{\sqrt{qE}}{\pi \beta }\tilde{n}^{cr}\;,
\nonumber \\
&<&T_{11}>^{(3)}=<T_{22}>^{(3)}=qH/\sqrt{qE}\sinh ^{-1}(2\pi H/E)\tilde{n}%
^{cr}\beta \ln \beta ^{-1}\;.  \label{gsf7}
\end{eqnarray}
We see that leading vacuum
polarization effects are  in main defined
 by the total 
density of excited vacuum states, which depend on the complete
 history of a state. Thus,
 it is a nonlocal contribution.

Considering the small time limit of $<T_{\mu \nu }>^{(2)}$ we get
\[
<T_{\mu \mu }>^{(2)}=l_\mu (s_2)\left. \mbox{tr}\left\{ \Delta
^{(2)}(x,x^{\prime })\right\} \right| _{x=x^{\prime }}\;,
\]
where $\mbox{tr} \Delta ^{(2)} $ was found in (%
\ref{App4}). If $\beta <<1$, the result has a  simple form: 
\begin{eqnarray*}
&<&T_{00}>^{(2)}=<T_{33}>^{(2)}=qEx_0\tilde{n}^{cr}\;, \\
&<&T_{11}>^{(2)}=<T_{22}>^{(2)}=-2qH\sinh ^{-1}(2\pi H/E)x_0\pi \beta ^2%
\tilde{n}^{cr}\;.
\end{eqnarray*}
One can see that expressions for
 $<T_{\mu \nu }>^{(2)}$ at the asymptotic time and at the small time
are quite different. The only components $<T_{00}>^{(2)}=<T_{33}>^{(2)}$ for 
$\beta <<1$ coincide.

The technics developed in the article allows one 
take into account  global effects of
both  a real particles creation and a polarization
on the same footing.
The explicit form and the limit estimations for the mean
vacuum values of EMT may be used
 for the calculations 
of the back reaction from the unstable vacuum to the gravitational
background (in particular computing simulation similar to \cite{HH},
see also \cite{HMM} and references therein).  We see 
 a behaviour of such components in time are quite different, and one needs 
to take into account characteristic polarization effect.  
The  proper-time representations of the GF 
may be the necessary step in
the study of chiral symmetry breaking in QED and the four-fermion
models under the action of gravitational and electromagnetic fields
 \cite{IMO}.
Such a
study may have an immediate important application to EU, for
example, through the construction of inflationary Universe where role of
inflaton is played by the condensate $<\bar{\Psi}\Psi>$. One can also
analyse symmetry breaking phenomenon under the combined action of
these
 fields in the Standard Model (using also its gauged NJL
form \cite{NJL}), or GUT in the same way as it has been
done in curved spacetime (without electromagnetic field) \cite{BOS}.
Our methods maybe applied to the study of SUSY NJL model
in curved spacetime introduced in \cite{BIO}. 
The chiral symmetry breaking under the action of weak gravitational
field 
and constant electromagnetic field has been studied in \cite{IOSh}. 
Having our results we may study the chiral symmetry breaking 
for such model in FRW Universe with constant electromagnetic field.


\section{Acknowledgments}

D.M.G. thanks Brazilian foundations CNPq and FAPESP
for support. S.P.G. thanks
Brazilian foundation CAPES for support, and the Department of Physics of UEL
and Department of Mathematical Physics of USP for hospitality.
We also thank Prof. S.D. Odintsov for useful discussions.

\appendix

\section{Small time expansion of $\Delta ^{(2)}$-function}

To calculate  small time expansion of $\mbox{tr}\left\{ \Delta
^{(2)}\right\} $  given by Eq. (\ref{a42a}), in the case 
$x_0^2<<\rho /(bM)^2$ and $x\rightarrow x^{\prime }$, one can use the 
method presented in App. B of Ref. \cite{GGO}. The
final result is  
\begin{eqnarray}
&&\mbox{tr}\left\{ \Delta ^{(2)}(x,x^{\prime })\right\} =e^{iq\Lambda
}\left\{ \left[ i(x_0+x_0^{\prime })c_2(bM)^2/\rho -qEy^3\right]
n^{(2)}/(qE)\varphi _0\right.   \nonumber \\
&&+\left. i\left[ -(x_0+x_0^{\prime })^3\frac{(bM)^4}{12qE\rho ^2}%
K(2)+(1/2)(x_0+x_0^{\prime })(y_3)^2qEK(0)\right] \right\} \,\;,  \label{App4}
\\
&&\varphi _0=\exp \left( i\frac{qE}2(x_0+x_0^{\prime })y^3-\frac \rho {4c_2}%
\left( \frac{qE}{bM}\right) ^2(x_0-x_0^{\prime })^2-\frac{\rho ^3(y_3)^2}{%
4c_2(bM)^2}\right)   \nonumber \\
&&\times
\exp \left( \frac 14y_{\perp }qF\cot (\pi qF/\rho )y_{\perp }\right) \;, 
\nonumber \\
&&K(l)=-\frac{c_2^{l+1}\left( \rho /(bM)\right) ^2}{c_2^2+\left( \frac{qE}{bM%
}\right) ^2+\left( \frac{qE}{bM}\right) ^4}\left\{ a^2M^2/\rho
-c_2^{-1}(l-1/2)-c_2^{-1}\left( \frac{qE}{bM}\right) ^2\right.   \nonumber \\
&&\left. -c_2\left( \frac{bM}{qE}\right) ^2+2qH/\rho \sinh ^{-1}(2c_2qH/\rho
)+2c_2^{-1}\frac{\left[ 1+\left( \frac{qE}{bM}\right) ^2\right] \left[
c_2^2+\left( \frac{qE}{bM}\right) ^4\right] }{c_2^2+\left( \frac{qE}{bM}%
\right) ^2+\left( \frac{qE}{bM}\right) ^4}\right\} n^{(2)}\;,  \nonumber \\
&&n^{(2)}=-1/2\tau (s_2)n_{sc}^{(2)}\;,  \nonumber \\
&&n_{sc}^{(2)}=\frac{\sqrt{c_2^2+\left( \frac{qE}{bM}\right) ^4}}{8\pi ^{3/2}%
\sqrt{c_2}\left[ c_2^2+\left( \frac{qE}{bM}\right) ^2+\left( \frac{qE}{bM}%
\right) ^4\right] }\frac{q^2HE\rho ^{5/2}}{(bM)^3\sin (c_2qH/\rho )}%
e^{-c_2a^2M^2/\rho }\;,  \nonumber
\end{eqnarray}
If $bM/qE<<1$, 
the coefficients $K(l)$ and $n^{(2)}$ from (\ref{App4}) have more
simple form. Thus, in the case of an intensive electric field ($%
a^2M^2/(qE)<1,\;|H/E|<1$) we get
\begin{equation}
n^{(2)}=\tilde{n}^{cr}\;,\;\;K(l)=-\pi ^l\tilde{n}^{cr}\;.
\end{equation}
Since the final formula (B4) in App. B of Ref.\cite{GGO}, which
 describes  small
time expansion of $\Delta _{sc}^{(2)}$ in scalar case,
was written with some misprints, we would like to
represent its corrected form here:

\begin{eqnarray*}
&&\Delta _{sc}^{(2)}(x,x^{\prime })=e^{iq\Lambda }\left\{ \left[
i(x_0+x_0^{\prime })c_2(bM)^2/\rho -qEy^3\right]
(-1/2)n_{sc}^{(2)}/(qE)\varphi _0\right. \\
&&+\left. i\left[ -(x_0+x_0^{\prime })^3\frac{(bM)^4}{12qE\rho ^2}%
K_{sc}(2)+(1/2)(x_0+x_0^{\prime })(y_3)^2qEK_{sc}(0)\right] \right\} \,\;, \\
&&K_{sc}(l)=-\frac{c_2^{l+1}\left( \rho /(bM)\right) ^2}{c_2^2+\left( \frac{%
qE}{bM}\right) ^2+\left( \frac{qE}{bM}\right) ^4}\left\{ a^2M^2/\rho
-c_2^{-1}(l-1/2)-c_2^{-1}\left( \frac{qE}{bM}\right) ^2\right. \\
&&\left. +(qH/\rho )\coth (c_2qH/\rho )+2c_2^{-1}\frac{\left[ 1+\left( \frac{%
qE}{bM}\right) ^2\right] \left[ c_2^2+\left( \frac{qE}{bM}\right) ^4\right] 
}{c_2^2+\left( \frac{qE}{bM}\right) ^2+\left( \frac{qE}{bM}\right) ^4}%
\right\} n_{sc}^{(2)}\;.
\end{eqnarray*}

\end{document}